\documentclass[reprint, secnumarabic,amssymb, superscriptaddress, nobibnotes, aps, prx]{revtex4-2}
\usepackage{amsmath,amssymb,amsfonts}
\usepackage{braket}
\usepackage{xfrac}
\usepackage{graphicx}
\usepackage{subfigure}
\usepackage{color}
\usepackage{pict2e}
\usepackage{enumitem}
\usepackage{multirow}
\usepackage{cancel}
\usepackage[hyphens]{xurl}
\usepackage[hidelinks]{hyperref}

\begin{document}

\title{ShaNQar: Simulator of Network Quantique}%

\author{Anand Choudhary}%
\email{anand\_c@ph.iitr.ac.in}
\affiliation{Department of Physics, Indian Institute of Technology (IIT) Roorkee, Roorkee - 247667, Uttarakhand, India}
\affiliation{School of Computer and Communication Sciences, École Polytechnique Fédérale de Lausanne (EPFL), 1015 - Lausanne, Switzerland}
\author{Ajay Wasan}
\email{awasan@ph.iitr.ac.in}
\affiliation{Department of Physics, Indian Institute of Technology (IIT) Roorkee, Roorkee - 247667, Uttarakhand, India}
\date{March 2025}%

\begin{abstract}
    The nature-inspired field of quantum communication has witnessed exciting developments over the past few years with countries all over the world working hard to scale their experimental quantum networks to larger sizes and increased coverage. Evidently, quantum network simulators are the need of the hour as they provide a framework for tuning hardware parameters, optimizing control protocols, and testing configurations of large and complex quantum networks before their deployment in the real world. In this work, we present ShaNQar (Simulator of Network Quantique): a modular and customizable photonic quantum network simulator. It comprises models of components such as photons, lasers, neutral density filters,  sources of entangled photon pairs, communication channels, mirrors, waveplates, beam splitters, single photon detectors, and nodes which incorporate a diverse set of tunable parameters for variability and versatility. It enables adaptive timing control and synchronization with virtually no simulation time resolution limit and features a `plug and play' design for faster coding and efficient execution. We successfully simulated previous real-life experimental setups for Quantum Key Distribution (QKD) and quantum teleportation, thereby, demonstrating the reliability and accuracy of ShaNQar. 
\end{abstract}

\maketitle

\section{Introduction}
\label{intro}

Secure and fast communication have become the \textit{sine qua non} in today's data-driven world. Quantum Communication promises to deliver on this very need of the hour by harnessing the quantum advantage which draws inspiration from the unrivalled computational and information processing capabilities of Nature and the Human Body. Features such as superposition, entanglement, \textit{true} randomness, and the no-cloning theorem impart quantum communication an edge over classical communication by lending, \textit{inter alia}, quantum communication networks the ability to compute exponentially faster and provide uncompromisable security to all their users~\cite{b1}.

Progress in the experimental realization of quantum networks has gained huge momentum with numerous operational Quantum Key Distribution (QKD) networks being deployed across the globe. Netherlands is at the forefront of quantum networking, having successfully developed the world's first entanglement-based quantum network~\cite{b2}. The US has also recently unveiled a 124 mile long quantum network connecting Chicago with the nearby suburban areas and Toshiba has already started testing the security of their QKD protocols on the network~\cite{b3}. Other countries such as Japan~\cite{b4}, China~\cite{b5}, South Korea~\cite{b6}, UK~\cite{b7} and Switzerland~\cite{b8} are also in the race with a focus on increasing the coverage of their quantum networks. India is also building an indigenous Quantum Internet with Local Access (QuILA) linking four Metro Access Quantum Area Networks (MAQANs): Delhi, Hyderabad, Bengaluru, and Chennai~\cite{b9}.

To increase the size, coverage, and complexity of these quantum networks which are spread out across the globe, quantum hardware development must be accelerated to newer heights and the associated control protocols must be made more robust and secure. Optimization of the resources associated with quantum networks and standardization of their architectures is also the need of the hour~\cite{b10}. There is thus a worldwide need for the development of platforms that can reliably and accurately simulate the actual quantum hardware and effectively model highly complex interconnections between them before such large quantum networks are implemented in actuality. Such simulation platforms must also be equipped with the capability to run various protocols associated with quantum communication and to test their scale and efficiency in complicated architectures with varying device and linkage parameters. Moreover, the simulation platform must be modular to allow for the building of highly complex networks from simple building blocks. It must also be highly customizable and reprogrammable, so as to allow for the modelling of \textit{promising} new components and links as and when they get developed. 

A coherent amalgamation of all the aforementioned concepts, developments, and ideas naturally gave rise to our main aim of building a quantum network simulator, fulfilling which, we here present \textbf{ShaNQar: Simulator of Network Quantique}(\textit{Quantique} is the French equivalent of \textit{Quantum}). All our code is available at the following link: \url{https://github.com/andy-joy-25/ShaNQar}.

The statue of Lord Shiva (Lord ShaNQar) in his Nataraja form stands tall at CERN, the European Center for Research in Particle Physics, in Geneva. This statue which symbolizes Shiva's cosmic dance of creation and destruction actually conveys the profound significance of the metaphor of Shiva's dance for the cosmic dance of subatomic particles (in the quantum mechanical regime), which is analyzed and observed by the physicists at CERN~\cite{b11}. This inspired us to name our Quantum Network Simulator as ShaNQar.

ShaNQar is a photonic quantum network simulator. We specifically chose the photonic modality for the purpose of simulation since the usage of photons as qubits allows for high speed, low loss, and controllable quantum communication~\cite{b1}. 

\section{ShaNQar's Hardware Stack}
\label{hardware_stack}

The hardware stack of our quantum network simulator, ShaNQar, features realistic models of practically useful physical components of quantum networks. Timing control and synchronization is enabled via the python based discrete event simulation framework, SimPy~\cite{b12}. All components, protocols, etc. thus share a simulation environment which effectively tracks and controls the simulation time and helps in executing a set of events in sequence. 

Since photons are essentially the fundamental units of quantum computing and quantum information processing (here), it is imperative that they be modelled separately as well. Additionally, the quantum state of a photon must be tracked and/or controlled right from that point in time when the photon is emitted and all the way up to the point when it finally gets detected. This clearly justifies the need for a separate model to accomplish the same.

We now describe all the models of physical components (including the models for the photon and its quantum state) in detail. All the component models and the model for the photon have a unique ID and an environment (which may be \textit{adaptive} (see Section~\ref{QTimpl} for an example which illustrates the usage of this feature)) which enables their individual tracking and control. 

\subsection{Quantum State}
\label{qstate}

The \textbf{QuantumState} class facilitates the accurate realism, tracking, manipulation, and/or control of the quantum state of photons (here, the qubits). Currently, we only support the polarization based encoding of quantum information. Thus, effectively, the polarization state of a photon defines its quantum state. We make use of the Jones Calculus so as to represent the state of polarization of a photon as a Jones vector. Following are the 3 sets of commonly used mutually orthogonal polarization states and their correspondence to the bases used in quantum computation and quantum information~\cite{b1,b13}:
\begin{itemize}
    \item \textbf{Horizontal $\ket{H}$ and Vertical $\ket{V}$ Polarizations: Z Basis}\\ \begin{equation} \ket{H} (\equiv \ket{0}) = \begin{bmatrix} 1\\0 \end{bmatrix}\label{eqna1i}\end{equation}\\\text{and}\\\begin{equation} \ket{V} (\equiv \ket{1}) = \begin{bmatrix} 0\\1 \end{bmatrix}. \label{eqna1ii} \end{equation}
    \item \textbf{Diagonal $\ket{D}$ and Anti-Diagonal $\ket{A}$ Polarizations: X Basis}\\ \begin{equation} \ket{D} (\equiv \ket{+}) = \frac{1}{\sqrt{2}}\!\begin{bmatrix} 1\\1 \end{bmatrix}\label{eqna2i}\end{equation}\\\text{and}\\\begin{equation} \ket{A} (\equiv \ket{-}) = \frac{1}{\sqrt{2}}\!\begin{bmatrix} 1\\-1 \end{bmatrix}. \label{eqna2ii} \end{equation}
    \item \textbf{Right Circular $\ket{R}$ and Left Circular $\ket{L}$ Polarizations: Y Basis}\\ \begin{equation} \ket{R} (\equiv \ket{+i}) = \frac{1}{\sqrt{2}}\!\begin{bmatrix} 1\\i \end{bmatrix}\label{eqna3i}\end{equation}\\\text{and}\\\begin{equation} \ket{L} (\equiv \ket{-i}) = \frac{1}{\sqrt{2}}\!\begin{bmatrix} 1\\-i \end{bmatrix}. \label{eqna3ii}\end{equation}
\end{itemize}
Any arbitrary polarization state of a photon $\ket{\psi}$ can clearly be written as a superposition of the aforementioned orthogonal polarization states corresponding to a particular basis, say, $\ket{\phi_1}$ and $\ket{\phi_2}$, such that $\ket{\psi} = a\!\ket{\phi_1} + b\!\ket{\phi_2}$ where, $a$ and $b$ represent the coefficients of the quantum state. $|a|^2$ and $|b|^2$ represent the probabilities of the photon being in the polarization states $\ket{\phi_1}$ and $\ket{\phi_2}$ respectively and hence, $|a|^2 + |b|^2 = 1$. In our simulator, a \textbf{QuantumState} object thus, has the coefficients as well as the corresponding basis as its attributes which serve to provide a complete realization of the quantum state.

Currently, we support quantum states of up to 2 qubits (single qubit states, product states, and entangled states). Product states are mathematically generated by taking the tensor product of 2 single qubit states: $\ket{\psi_{prod}} = \ket{\psi_{s1}} \otimes \ket{\psi_{s2}}$.

Apart from the (aforementioned) ket formalism of the quantum state, we also support the density matrix formalism~\cite{b1}. The density matrix corresponding to a pure quantum state $\ket{\psi}$ is given by $\rho = \ket{\psi}\!\bra{\psi}$. For the more general case of a quantum system that can be found in any state $\ket{\psi_i}$ (with a probability $p_i$) out of an ensemble of pure states $\{p_i,\ket{\psi_i}\}$, the density matrix is defined by $\rho = \sum_i p_i \ket{\psi_i}\!\bra{\psi_i}$. 

Further, we also provide the functionality for interconversion between the 2 formalisms. While the conversion from the ket to the density matrix formalism follows from the formulae mentioned above, the conversion from the density matrix to the ket formalism is not so straightforward. This is because it quite possible for a number of ensembles of pure states to give rise to the same density matrix. Accordingly, it's only practical to select the state vector (eigenvector) with the highest corresponding probability (eigenvalue) in the spectral decomposition of the density matrix while affecting the said conversion~\cite{b131}. 

Since no physical system can possibly be completely noise free, quantum states of photons are inevitably prone to get corrupted (partially or fully) by noise at any and all stages of the quantum network. In order to account for the same, we model the following different types of dissipative and non-dissipative noise~\cite{b1} (see Fig.~\ref{noise} for an overview):

\begin{figure}[htbp]
\begin{center}
\includegraphics[scale = 0.50]{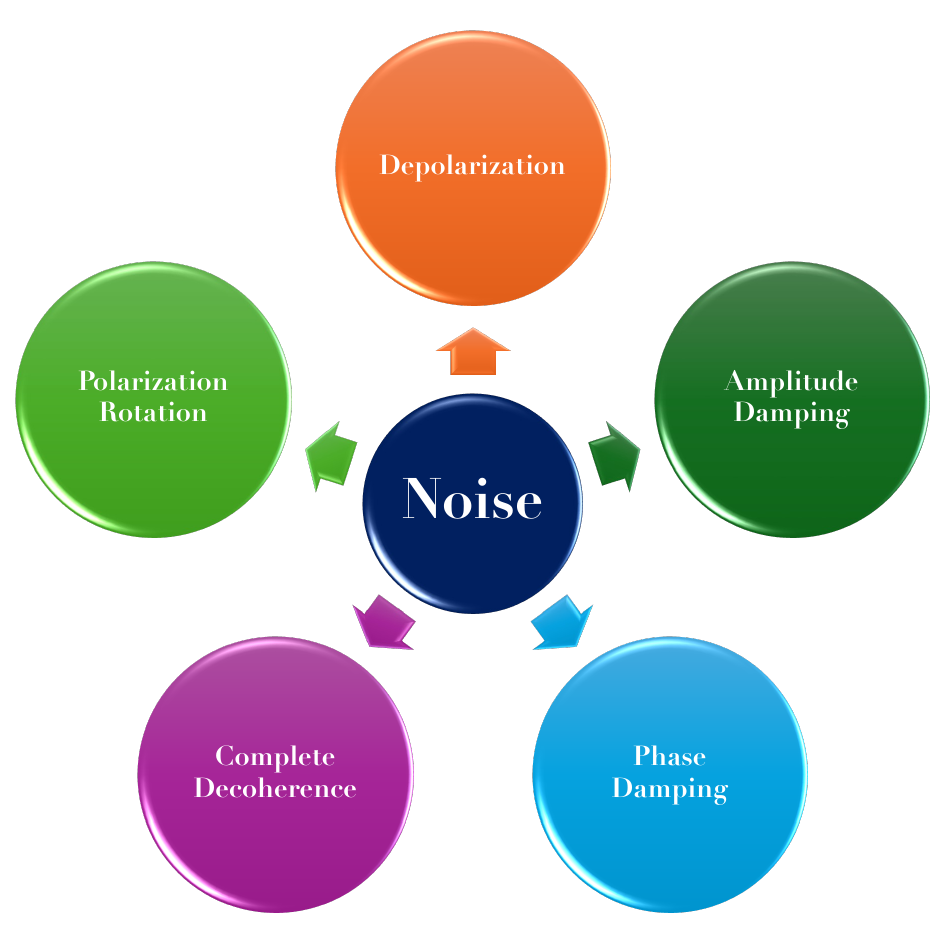}
\end{center}
\caption{Different types of noise modelled within the QuantumState class in ShaNQar}
\label{noise}
\end{figure}

\begin{itemize}
    \item \textbf{Polarization Rotation Noise}: It rotates the polarization state of a photon by a random angle $\theta \in [0, 2\pi]$ about an axis which forms the basis of its polarization state. The following rotation operators are used for affecting the said effect for the case of single qubit states:
    \begin{equation}
        R_x(\theta) \equiv e^{\frac{-i\theta X}{2}} = \begin{bmatrix} \cos{\frac{\theta}{2}} & -i \sin{\frac{\theta}{2}}\\-i \sin{\frac{\theta}{2}} & \cos{\frac{\theta}{2}}\end{bmatrix},
     \label{eqna4}
     \end{equation}
     \begin{equation}
        R_y(\theta) \equiv e^{\frac{-i\theta Y}{2}} = \begin{bmatrix} \cos{\frac{\theta}{2}} & -\sin{\frac{\theta}{2}}\\\sin{\frac{\theta}{2}} & \cos{\frac{\theta}{2}}\end{bmatrix},
    \label{eqna5}
    \end{equation}
    and
     \begin{equation}
        R_z(\theta) \equiv e^{\frac{-i\theta Z}{2}} = \begin{bmatrix} e^{\frac{-i\theta}{2}} & 0\\0 & e^{\frac{i\theta}{2}}\end{bmatrix}.
    \label{eqna6}
    \end{equation}
    Here, $X$, $Y$, and $Z$ are the Pauli matrices.
    
    For 2 qubit states (product or entangled states), tensor products of the aforementioned rotation operators are used: $R_x(\theta) \otimes R_x(\theta), R_y(\theta) \otimes R_y(\theta)$, and $R_z(\theta) \otimes R_z(\theta)$ so as to rotate the polarization states of both the qubits by the same amount.
    \item \textbf{Depolarization Noise}: In line with the quantum operations formalism, the density matrix (for a d-dimensional quantum system) transforms as:
    \begin{equation}
        \rho \rightarrow \mathcal{E}_P(\rho) = \frac{pI}{d} + (1 - p) \rho.
    \label{eqna7}
    \end{equation}
    Physically, what this means is that the depolarization noise depolarizes the quantum state, i.e., replaces it with the completely mixed state $\frac{I}{d}$, with a probability = $p$ and does not affect it at all with a probability = $1 - p$. 
    \item \textbf{Amplitude Damping Noise}: This models the loss of energy to the environment. Effectively, in the quantum operations formalism, the density matrix $\rho$ corresponding to a single qubit state $\ket{\psi}$ transforms as:
    \begin{equation}
        \rho \rightarrow \mathcal{E}_{AD}(\rho) = E_0 \rho E_0^{\dag} + E_1 \rho E_1^{\dag},
    \label{eqna8}
    \end{equation}
    where, 
    \begin{equation}
        E_0 = \begin{bmatrix} 1 & 0\\0 & \sqrt{1 - \gamma}\end{bmatrix},
    \label{eqna9}
    \end{equation}
    \begin{equation}
        E_1 = \begin{bmatrix} 1 & \sqrt{\gamma}\\0 & 0\end{bmatrix},
    \label{eqna10}
    \end{equation}
    and $\gamma$ is the probability of losing a photon. 
    
    For 2 qubit states, the quantum operation $\mathcal{E}_{AD} \otimes \mathcal{E}_{AD}$ is used to affect the transformation of their density matrices under the effect of amplitude damping.
    \item \textbf{Phase Damping Noise}: Also referred to as the dephasing noise, this models the loss of quantum information without any energy loss. Physically, this leads to the accumulation of a random relative phase between the eigenstates of a quantum system. In the quantum operations formalism, the density matrix $\rho$ corresponding to a single qubit state $\ket{\psi}$ transforms as:
    \begin{equation}
        \rho \rightarrow \mathcal{E}_{PD}(\rho) = E_0 \rho E_0^{\dag} + E_1 \rho E_1^{\dag},
    \label{eqna11}
    \end{equation}
    where, 
    \begin{equation}
        E_0 = \begin{bmatrix} 1 & 0\\0 & \sqrt{1 - \lambda}\end{bmatrix},
    \label{eqna12}
    \end{equation}
    \begin{equation}
        E_1 = \begin{bmatrix} 0 & 0\\0 & \sqrt{\lambda}\end{bmatrix},
    \label{eqna13}
    \end{equation}
    and $\lambda$ is the probability of the lossless scattering of a photon. 
    
    For 2 qubit states, the quantum operation $\mathcal{E}_{PD} \otimes \mathcal{E}_{PD}$ is used to affect the transformation of their density matrices under the effect of phase damping.
    \item \textbf{Complete Decoherence Noise}: This combines the effect of both amplitude and phase damping noise and hence, affects the following transformation of the density matrix $\rho$~\cite{b14}:
    \begin{equation}
        \rho \rightarrow \mathcal{E}_{DC}(\rho) \equiv \mathcal{E}_{PD} \circ \mathcal{E}_{AD} (\rho) \equiv \mathcal{E}_{PD}(\mathcal{E}_{AD}(\rho)).
    \label{eqna14}
    \end{equation}
    where, $\mathcal{E}_{PD}$ and $\mathcal{E}_{AD}$ are the quantum operators corresponding to the phase damping noise and amplitude damping noise respectively, as defined in (\ref{eqna8}) and (\ref{eqna11}).
\end{itemize}

Now, it's inevitably necessary to measure the state of the photon(s) transmitted by the sender node at the receiver node. Mathematically (we describe the actual physical implementation of quantum state measurement later), the quantum state of photons is measured by means of a measurement operator, say, $M_m$. We implement the measurement of a quantum state based on the fact that the probability of obtaining an outcome $m$ upon affecting the measurement of a quantum state using $M_m$ is given by~\cite{b1}:
\begin{equation}
    p(m) = \braket{\psi|M_m^\dag M_m|\psi},
    \label{eqna15}
\end{equation}
and the post-measurement quantum state is given by:
\begin{equation}
    \ket{\psi '} = \frac{M_m \ket{\psi}}{\sqrt{p(m)}}.
    \label{eqna16}
\end{equation}

Further, $M_m$ belongs to a set $M$ of measurement operators which satisfy the completeness relation: $\sum_m M_m^{\dag} M_m = I$, a reexpression of the fact that $\sum_m p(m) = 1$. Measurement is commonly performed using the projectors $P_m = \ket{m}\!\bra{m}$ to effectively measure the quantum state $\ket{\psi}$ in a basis $\ket{m}$.

Finally, after measurement, let's say that we obtain the final state as $\ket{\psi '} \equiv \mathcal{E}(\ket{\psi}\!\bra{\psi})$ where $\mathcal{E}$ is the quantum operation corresponding to a physical component/system governing the transformation of the quantum states of the photons when the photons pass through it. It's evident that $\ket{\psi '} \neq \ket{\psi}$ given that noise (apart from any other possible external sources of error such as eavesdropping, etc.) does indeed corrupt the quantum states of photons by changing an initial quantum state $\ket{\psi}$ to the final state $\mathcal{E}(\ket{\psi}\!\bra{\psi})$. It's however, worth investigating the amount by which it does the same so as to ascertain the amount of error introduced during transmission of photons through a physical component/system and hence, for instance, identify the means to decrease that error as much as possible. Alternatively said, it's worthwhile to figure out the extent to which any given physical component/system is able to preserve the state $\ket{\psi}$ of the system by comparing the extent of the similarity between $\ket{\psi}$ and $\mathcal{E}(\ket{\psi}\!\bra{\psi})$. We accomplish this by implementing functions for the computation of two commonly used measures for quantum information: the trace distance and the fidelity~\cite{b1}, which can be invoked by the user wherever required. 

The trace distance between 2 quantum states is defined in terms of their respective density matrices $\rho$ and $\sigma$ as follows:
\begin{equation}
    D(\rho, \sigma) = \frac{1}{2} \text{tr} (|\rho - \sigma|),
    \label{eqna17}
\end{equation}
where, $|P| \equiv +\,\sqrt{P P^\dag}$.

On the other hand, the fidelity between $\rho$ and $\sigma$ is defined as follows:
\begin{equation}
    F(\rho, \sigma) = \text{tr} \left(\sqrt{\rho^{\frac{1}{2}} \sigma \rho^{\frac{1}{2}}}\right).
    \label{eqna18}
\end{equation}

Both $D(\rho, \sigma)$ and $F(\rho, \sigma) \in [0,1]$, but while the indistinguishability between 2 quantum states is implied by a trace distance of zero, the same is implied by a fidelity of one. It turns out that the trace distance and the fidelity are \textit{qualitatively} equivalent measures of similarity between quantum states. The usage of the fidelity is somewhat preferred over the trace distance, however, since it turns out to be easier to work with. 

\subsection{Photon}
\label{photon}

The \textbf{Photon} class is a model of a photon, the quantum of light. Each \textbf{Photon} object possesses a wavelength, linewidth, temporal width, an associated quantum state, and the corresponding scheme of quantum information encoding (currently, only polarization) as its attributes. The photon's finite linewidth, is effectively, the linewidth of the quasi-monochromatic source, for e.g., a laser, which emitted it in the first place in the quantum network. The temporal width of the photon is defined as the standard deviation of the probability distribution function for the time of its arrival at the destination point~\cite{b15}. The temporal width is an important attribute which captures the effect of dispersion accrued during its propagation through a quantum channel. 

\subsection{Laser}
\label{laser}

A laser is a device which emits light via the process of feedback amplification of light by the stimulated emission of radiation~\cite{b16}. It emits highly monochromatic, directional, and coherent light and is thus used as a preferred source of photons for long distance classical as well as (after attenuation to single photon levels) quantum communication networks. The laser outputs a coherent state $\ket{\alpha}$ given by~\cite{b1}:
\begin{equation}
    \ket{\alpha} = \sum_{n = 0}^\infty e^{-\frac{| \alpha |^2}{2}} \frac{\alpha^n}{\sqrt{n!}}\!\ket{n},
    \label{eqna19}
\end{equation}
where $\ket{n}$ is the n-photon energy eigenstate. The number of photons emitted by a laser thus follows a poisson distribution with mean = $\alpha^2$. 

In our simulator, the \textbf{Laser} class models a pulsed laser which emits a random number of photons drawn from a poisson distribution with a user specified mean number of photons = $\mu_{p}$ ($\equiv \alpha^2$) in every pulse. The time period of each pulse is decided by the repetition rate of the laser: PRR (another user configurable attribute). Some of the other common attributes include the wavelength $\lambda$, linewidth $\Delta \lambda$ and temporal width $W^L_{temp}$.

The Laser class also implicitly models a polarizer by emitting the photons in user specified quantum state(s). In order to account for the non-ideal behaviour of the polarizer, we have included the polarization extinction ratio (PER) of the polarizer as a user specified attribute. The PER is defined as the transmission ratio of the wanted to the unwanted component(s) of polarization and based on its value, few of the emitted photons (chosen at random) can be emitted in the unwanted polarization state. Non-ideality of the laser has also been considered in our model since the quantum states of a few of the emitted photons (again chosen at random) can get corrupted by complete decoherence noise~(\ref{eqna14}) depending upon the user specified noise level (NL) of the laser. 

\subsection{Neutral Density Filter}
\label{NDfilter}

In a quantum communication network, one deals with the transmission, detection, and control of the quantum states of individual photons and hence, it is necessary to attenuate the output of a laser to single photon levels. An absorptive Neutral Density (ND) filter uniformly attenuates light over a significant wavelength range by absorbing a fraction of it and is often used to attenuate the output of lasers to a desired level. The extent to which an absorptive ND filter attenuates a laser's output is quantified in terms of its absorbance or, equivalently, its Optical Density ($OD$). Accordingly, the transmittance ($T$) of an ND filter is given by:
\begin{equation}
    T = 10^{-OD}.
    \label{eqna20}
\end{equation}

The \textbf{NDFilter} class in our simulator models a variable ND filter which essentially allows for setting the $OD$ of the filter to any desired value between user defined minimum and maximum $OD$ values. This is of great advantage since a greater tunability allows the selection of an appropriate $OD$ to achieve a desired attenuation corresponding to the mean number of photons that we finally want the (thus made) weak laser to emit. 

Our model of the ND filter enforces allowable yet (at the same time) random transmission of photons. Effectively, what this means is that the energy of the photon(s) transmitted by the ND filter is less than or equal to the maximum allowable value of the transmission energy of the ND filter (as determined by its transmittance) and photons are randomly transmitted based on whether their probability of being transmitted is less than the transmittance of the ND filter or not. Hence, the modelling of the ND filter has been done in line with both classical (a deterministic amount of transmitted energy) as well as quantum mechanical (a probabilistic emission of photons) viewpoints. 

\subsection{Weak Laser}
\label{weaklaser}

The \textbf{Weaklaser} class models a weak laser by inheriting the \textbf{Laser} and \textbf{NDFilter} classes. Effectively, it makes the job of the user easier by inherently selecting an appropriate number of ND filter(s) ($N_{NDF}$) and their corresponding $OD$(s) so as to weaken, i.e., attenuate the output of the laser and hence, yield single photon pulses (on average). 

Naturally, it's desirable to minimize the number of ND filters required for the said attenuation. Consequently, at first, we check the value of the net $OD$ required for the same. This value is decided on the basis of the mean number of photons which are initially being emitted by the laser. So, if the laser initially emits $\mu_i$ number of photons (on average) and if after attenuation, we wish to ensure that it only emits $1$ photon on average, the (average) net transmittance of the corresponding ND filter must be:
\begin{equation}
    T_{net} = \frac{1}{\mu_i},
    \label{eqna21}
\end{equation}
and hence, its net $OD$ must be:
\begin{equation}
    OD_{net} = \log_{10}\left(\frac{1}{T_{net}}\right) = \log_{10} (\mu_i).
    \label{eqna22}
\end{equation}
If $OD_{net} < $ maximum $OD$ of a single ND filter, then only a single ND filter can do the job. Otherwise, a stack of ND filters is used in order to achieve the $OD_{net}$ value up to a pre-specified level of tolerance. The stack of ND filters is preferably (and only if possible) chosen in such a manner that the $OD$ of each ND filter is the same. 

\subsection{Source of Entangled Photon Pairs}
\label{SPDC}

A composite quantum system is said to be in an entangled state if its quantum state is not expressible as a tensor product of the states of its individual components. Physically, this means that it is impossible to completely describe the state of any of its components in isolation from the others. Essentially, components share an inextricable bond with one another irrespective of how far they are located from one another. Measurements of the states of the components will thus necessarily be correlated with one another. A perfect correlation implies maximal entanglement. This is the case for the Bell states/EPR(Einstein, Podolsky and Rosen) states/EPR pairs~\cite{b17,b18}. The Bell states are given by~\cite{b1}:
\begin{equation}
    \ket{\phi^{\pm}} = \frac{1}{\sqrt{2}} (\ket{00} \pm \ket{11}), 
    \label{eqna23}
\end{equation}
and
\begin{equation}
    \ket{\psi^{\pm}} = \frac{1}{\sqrt{2}} (\ket{01} \pm \ket{10}).
    \label{eqna24}
\end{equation}
An imperfect, i.e., a partial correlation, on the other hand, implies non-maximal entanglement. Non-maximally entangled states are of the form~\cite{b20}:
\begin{equation}
    \ket{\widetilde{\phi}^{\pm}} = \frac{1}{\sqrt{1 + \epsilon^2}} (\ket{00} \pm \epsilon \ket{11}), 
    \label{eqna25}
\end{equation}
and
\begin{equation}
    \ket{\widetilde{\psi}^{\pm}} = \frac{1}{\sqrt{1 + \epsilon^2}} (\ket{01} \pm \epsilon \ket{10}).
    \label{eqna26}
\end{equation}
These states provide a controllable degree of entanglement ($\epsilon$) and while their applications in various domains of quantum communication are still in a nascent stage, they might very well prove to be a boon for futuristic QCQI applications~\cite{b20}. 

\begin{figure}[htbp]
\begin{center}
\includegraphics[scale = 0.25]{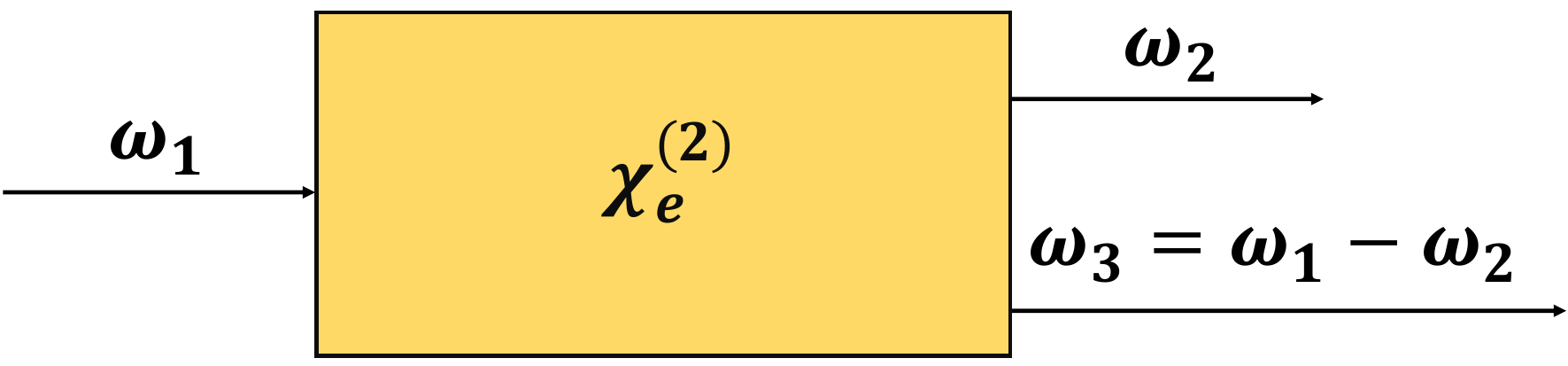}
\end{center}
\caption{Spontaneous Parametric Down Conversion (SPDC)}
\label{FigSPDC}
\end{figure}

In their seminal work in 1995, Kwiat \textit{et al}~\cite{b19} experimentally realised the production of polarization-entangled photon pairs by using a (frequency) degenerate Type-II Spontaneous Parametric Down-Conversion (SPDC) process involving non-collinear phase matching using a $\beta$-BBO (Barium Borate: Ba(BO\textsubscript{2})\textsubscript{2}) crystal. SPDC is a spontaneous 2 photon emission process wherein a pump photon at $\omega_1$ frequency is down converted to form a signal photon at $\omega_2$ frequency and an idler photon at $\omega_3$ frequency. It only occurs in non-centrosymmetric crystals, i.e., crystals with a non-zero second-order electrical susceptibility ( $\chi^{(2)}_e$ ) ~\cite{b21}. SPDC has been schematically depicted in Fig.~\ref{FigSPDC}. 

In~\cite{b19}, they found out that the down-converted photons are emitted into 2 cones (centered on the pump beam with their opening angle being determined by the angle $\theta_{pm}$ between the crystal’s optic axis and the pump) with one of them being ordinary polarized and the other one being extraordinary polarized. They further found out that increasing the value of $\theta_{pm}$ resulted in the 2 cones tilting towards the pump which in turn caused an intersection along 2 lines. The 2 overlap directions (labelled as 1 and 2) thus, represented the entangled state emission directions with the state of light being described by:
\begin{equation}
    \ket{\psi} = \frac{1}{\sqrt{2}} (\ket{H_1,V_2} + e^{i \alpha} \ket{V_1,H_2}).
    \label{eqna27}
\end{equation}
where, $H$ = Horizontal (Extraordinary) Polarization, $V$ = Vertical (Ordinary) Polarization, and $\alpha$ = (Extra) Relative Phase Factor arising from the birefringence of the BBO Crystal.  

With a rotation of the crystal, the value of $\alpha$ could be set as desired and a half-waveplate with $\theta = \frac{\pi}{4}$ (see (\ref{eqna35}) and~(\ref{eqna36})) in the path of one of the emitted photons was able to change the horizontal polarization to vertical polarization and vice versa. Consequently, they were able to obtain all the 4 maximally entangled Bell states as given in~(\ref{eqna23}) and~(\ref{eqna24}). 

In 1999, White \textit{et al}~\cite{b20} successfully generated non-maximally entangled states using Type-I Spontaneous Parametric Down Conversion (SPDC) with 2 thin $\beta$-BBO crystals instead of just one. Alignment of the 2 crystals was such that their optic axes were perpendicular to each other; the optic axis of the 1\textsuperscript{st} crystal defined the vertical plane and the optic axis of the 2\textsuperscript{nd} crystal defined the horizontal plane. Accordingly, with the pump polarization at an angle of $\frac{\pi}{4}$ with the vertical, down-conversion would be equally likely to occur in either crystal. Consequently, due to a high spatial overlap between the 2 processes of down-conversion (as discussed earlier), photons were emitted in the maximally entangled state given by:
\begin{equation}
    \ket{\psi} = \frac{1}{\sqrt{2}} (\ket{HH} + e^{i \alpha} \ket{VV}).
    \label{eqna28}
\end{equation}

For the generation of non-maximally entangled states, they simply rotated the pump polarization. When the pump polarization made an angle of $\chi$ w.r.t. the vertical, they obtained the following output entangled state:
\begin{equation}
    \ket{\psi} = \frac{1}{\sqrt{1 + \epsilon^2}} (\ket{HH} + \epsilon e^{i \alpha} \ket{VV}),
    \label{eqna29}
\end{equation}
where, $\epsilon = \tan(\chi)$.

As discussed earlier, an appropriate wave plate can be used to obtain states of the same form as that of $\ket{\widetilde{\phi}^{\pm}}$ (as defined in (\ref{eqna25})). Clearly, this experimental setup in~\cite{b20} is more general than the one in~\cite{b19} simply because it allows for the generation of both maximally as well as non-maximally entangled states. 

To allow for a user-configurable type of SPDC, however, the \textbf{EntangledPhotonsSourceSPDC} class in ShaNQar models the SPDC source given in~\cite{b19} and~\cite{b20} and allows for the configuration of the angle $\chi$ for Type-I SPDC~\cite{b20}. It further takes into account the user defined conversion efficiency of the SPDC process. This value of the conversion efficiency is generally very low, approximately of the order of 1 pair per $10^6$ pump photons~\cite{b21}. Hence, only a very few number of pump photons will be able to successfully down-convert and generate entangled photon pairs.

\subsection{Communication Channels}
\label{commchannels}

\subsubsection{Quantum Channel}
\label{qc}

The \textbf{QuantumChannel} class models a single mode fiber which is used for the transmission of photons (quantum information carriers). Its key (user-configurable) attributes include its length ($L$), the refractive index of its core ($n_{core}$), its attenuation coefficient ($\alpha$), chromatic dispersion coefficient ($D_{chr}$), polarization fidelity ($F_{pol}$) (the probability that a photon maintains its state of polarization while propagating through the fiber), depolarization probability ($p$) (as defined in (\ref{eqna7})), and coupling efficiency with the source ($\eta^{QC}_c$). 

Evidently, there is a finite amount of time delay in the arrival of photons at the receiving end of the quantum channel. The mean transmission time is clearly given by:
\begin{equation}
    t_{tr, mean} = \frac{L}{\left(\sfrac{c}{n_{core}}\right)},
    \label{eqna30}
\end{equation}
where, $c\,(= 3 \times 10^8 $ m/s) is the speed of light in vacuum.

As discussed earlier in Section~\ref{photon}, the actual transmission time varies from photon to photon and is dependent upon the temporal width $W^{QC}_{temp}$ acquired by a photon on passing through the channel. Clearly, 
\begin{equation}
    W^{QC}_{temp} = D_{chr}\Delta \lambda_{source}L,
    \label{eqna31}
\end{equation}
where, $\lambda_{source}$ = Wavelength of the source which emitted the photon.

Since $W^{QC}_{temp}$ is a standard deviation, the actual transmission time of a photon ($t_{tr, act}$) is given by:
\begin{equation}
    t_{tr, act} = t_{tr, mean} + W^{QC}_{temp}r_{norm},
    \label{eqna32}
\end{equation}
where, $r_{norm} \sim \mathcal{N}(0,\,1)$.
 
The number of photons that can possibly be coupled into the fiber depends upon the value of $\eta^{QC}_c$. Furthermore, transmission of photons that have been successfully coupled into the fiber is limited by the finite transmittance of the fiber which is given by:
\begin{equation}
    T = 10^{-\frac{\alpha L}{10}}.
    \label{eqna33}
\end{equation}

Successfully transmitted photons are further subject to being corrupted by noise with a probability  = $1 - F_{pol}$. Moreover, photons that are corrupted by noise are equally likely to be subjected to polarization rotation and depolarization noise (as described earlier in Section~\ref{qstate}). 

\subsubsection{Classical Channel}
\label{cc}

The \textbf{ClassicalChannel} models a separate optical fiber for the purpose of transmission of classical information. Transmission of information via a classical channel is assumed to be a lossless and completely reliable process, although it's subjected to a time delay which is the same as the mean transmission time given in (\ref{eqna30}). 

\subsection{Mirror}
\label{mirror}

The \textbf{Mirror} class models a mirror which reflects an incoming photon depending upon its reflectance. Additionally, the noise level of the mirror decides whether or not a reflected photon's quantum state would be corrupted by dissipative noise (see Section~\ref{qstate}).

\subsection{Waveplate}
\label{waveplate}

The \textbf{WavePlate} class models a general waveplate. An optical waveplate/retarder changes the polarization state of a photon passing through it by introducing a phase shift/retardance ($\alpha$) between the polarization components directed along its fast axis and slow axis by virtue of its birefringence (polarization dependence of the refractive index). Using Jones calculus, the generalised Jones matrix ($M_{WP}$) for a linear waveplate (i.e., one that introduces a linear phase retardance) is given by~\cite{waveplate_eqn_book}:
\begin{equation}
    M_{WP} = e^{-\frac{i \alpha}{2}} \begin{bmatrix} \cos^2{\theta} + e^{i \alpha} \sin^2{\theta} & (1 - e^{i \alpha}) \cos{\theta}\sin{\theta}\\ (1 - e^{i \alpha}) \cos{\theta}\sin{\theta} & \sin^2{\theta} + e^{i \alpha} \cos^2{\theta}\end{bmatrix}.
    \label{eqna34}
\end{equation}
where, $\theta$ = Orientation of the fast axis w.r.t. the horizontal, i.e, the $x$-axis.

For a half-waveplate, $\alpha = \pi$ and for a quarter-waveplate, $\alpha = \frac{\pi}{2}$. 

Let's examine the case of the half-waveplate in more detail. The Jones matrix for a half-waveplate ($M_{HWP}$) can clearly be obtained by substituting $\alpha = \pi$ in (\ref{eqna34}). Accordingly, we get that:
\begin{align}
    M_{HWP} & = e^{-\frac{i \pi}{2}} \begin{bmatrix} \cos^2{\theta} + e^{i \pi} \sin^2{\theta} & (1 - e^{i \pi}) \cos{\theta}\sin{\theta}\\ (1 - e^{i \pi}) \cos{\theta}\sin{\theta} & \sin^2{\theta} + e^{i \pi} \cos^2{\theta} \end{bmatrix} \\ \notag
    & = e^{-\frac{i \pi}{2}} \begin{bmatrix} \cos^2{\theta} - \sin^2{\theta} & 2\cos{\theta}\sin{\theta}\\ 2\cos{\theta}\sin{\theta} & \sin^2{\theta} - \cos^2{\theta} \end{bmatrix} \\ \notag
    & = e^{-\frac{i \pi}{2}} \begin{bmatrix} \cos{2\theta} & \sin{2\theta}\\ \sin{2\theta} & -\cos{2\theta} \end{bmatrix}. 
    \label{eqnab34}
\end{align}
Evidently, the global phase factor of $e^{-\frac{i \pi}{2}}$ is irrelevant as far as physical measurements are concerned and hence, can be ignored for further analysis here. 

Now, 2 cases are of special interest here. First, consider $\theta = \frac{\pi}{4}$. This yields:
\begin{equation}
    M_{HWP,\,\frac{\pi}{4}} = \begin{bmatrix} 0 & 1\\ 1 & 0 \end{bmatrix}. 
    \label{eqna35}
\end{equation}

Clearly, $M_{HWP,\,\frac{\pi}{4}} \equiv X$. $X$, as we know, is the quantum NOT gate, which is described by the action:
\begin{equation}
    X\!\ket{0} = \ket{1} \text{and}\,X\!\ket{1} = \ket{0}.
    \label{eqna36}
\end{equation}
As discussed in Section~\ref{SPDC}, a half-waveplate with $\theta = \frac{\pi}{4}$ can thus be used to switch between the horizontal and vertical polarization states.

Secondly, let's consider $\theta = \frac{\pi}{8}$. This yields:
\begin{equation}
    M_{HWP,\,\frac{\pi}{8}} = \frac{1}{\sqrt{2}} \begin{bmatrix} 1 & 1\\ 1 & -1 \end{bmatrix}. 
    \label{eqna37}
\end{equation}

Clearly, $M_{HWP,\,\frac{\pi}{8}} \equiv H$, where, $H$ denotes the Hadamard gate. $H$ is described by the action:
\begin{equation}
    H\!\ket{0} = \ket{+}, H\!\ket{1} = \ket{-}, H\!\ket{+} = \ket{0} \text{and}\,H\!\ket{-} = \ket{1}.
    \label{eqna38}
\end{equation}
Hence, a half-waveplate with $\theta = \frac{\pi}{8}$ can thus be used to switch from an orthogonal basis state in one basis ($Z$ or $X$) to the corresponding orthogonal basis state in the other basis ($X$ or $Z$, respectively).

Varying the values of $\theta$ and $\alpha$ can similarly yield some other different quantum gates. Waveplates thus function as quantum gates in a photonic quantum network.

\subsection{Beam Splitters}
\label{BS}

A beam splitter is an optical device used to split an incident beam of light into 2 beams. A beam splitter cube (2 triangular glass prisms glued together with optical cement) is preferred over a beam splitter plate (partially reflecting mirrors) for quantum communication applications since it is more durable and minimizes the transverse shift between the transmitted and incident beams~\cite{b22}. Further, since we operate in the (approximately) single photon regime in photonic quantum networks, the amount of power incident on the beam splitters is extremely low and hence, there is no possibility of damage of the cube. Hence, ShaNQar contains models of beam splitter cubes. 

The 2 types of beam splitters (for splitting the incident beam by intensity) are as follows:

\subsubsection{Non-Polarizing Beam Splitter (NPBS)} 
\label{NPBS}

While classically (for significantly high incident power levels involving a large number of photons) a non-polarizing beam splitter (NPBS) is described as reflecting a fraction $R$ of the incident light to one output port and transmitting the remaining fraction $1 - R$ to the other output port regardless of the polarization state of the light at the input, in the quantum mechanical regime (for extremely low power levels involving only $\approx$ a single photon), the reflectance and transmittance take the form of probabilities of reflection and transmission respectively. 

Considering the quantum mechanical regime, the transformation affected by a lossless, symmetric NPBS on the input field mode operators is given by~\cite{b23}:
\begin{equation}
\begin{bmatrix} \hat{a}_3 \\ \hat{a}_4\end{bmatrix} = \begin{bmatrix} t & r \\ r & t \end{bmatrix} \begin{bmatrix} \hat{a}_1 \\ \hat{a}_2 \end{bmatrix},
\label{eqnbs1}
\end{equation}
where, $\{\hat{a}_i\}_{i = 1\dots4}$ represent the annihilation operators corresponding to the input field modes (subscripts 1 and 2) and output field modes (subscripts 3 and 4) and $t$ and $r$ represent the transmittance and the reflectance of the NPBS respectively. 

Further, (here) $B$ denotes the transformation matrix of the NPBS such that:
\begin{equation*}
B = \begin{bmatrix} t & r \\ r & t \end{bmatrix}.
\end{equation*}

Since $B$ is unitary, $B B^\dag = I$ and thus,
\begin{equation*}
\begin{bmatrix} t & r \\ r & t \end{bmatrix} \begin{bmatrix} t^* & r^* \\ r^* & t^* \end{bmatrix} = \begin{bmatrix} 1 & 0 \\ 0 & 1 \end{bmatrix},
\end{equation*}

which upon simplification yields:
\begin{equation}
|t|^2 + |r|^2 = 1,
\label{eqnbs2}
\end{equation}
and
\begin{equation}
t r^* + r t^* = 0.
\label{eqnbs3}
\end{equation}

Clearly, the reverse NPBS transformation matrix (input in terms of output) can be easily found out by taking the adjoint of $B$. This then yields:
\begin{equation}
\begin{bmatrix} \hat{a}_1 \\ \hat{a}_2\end{bmatrix} = \begin{bmatrix} t^* & r^* \\ r^* & t^* \end{bmatrix} \begin{bmatrix} \hat{a}_3 \\ \hat{a}_4 \end{bmatrix}.
\label{eqnbs4}
\end{equation}

We can rewrite (\ref{eqnbs4}) as:
\begin{equation}
\hat{a}_1 = t^* \hat{a}_3 + r^* \hat{a}_4,
\label{eqnbs5}
\end{equation}
and
\begin{equation}
\hat{a}_2 = r^* \hat{a}_3 + t^* \hat{a}_4.
\label{eqnbs6}
\end{equation}

Now, since the creation operators ($\{\hat{a}^\dag_i\}_{i = 1\dots4}$) are the adjoints of the annihilation operators ($\{\hat{a}_i\}_{i = 1\dots4}$), the expressions relating the input creation operators to the output creation operators can easily be found by taking the adjoint of (\ref{eqnbs5}) and (\ref{eqnbs6}). We, thus get that:
\begin{equation}
\hat{a}^\dag_1 = t \hat{a}^\dag_3 + r \hat{a}^\dag_4,
\label{eqnbs7}
\end{equation}
and
\begin{equation}
\hat{a}^\dag_2 = r \hat{a}^\dag_3 + t \hat{a}^\dag_4.
\label{eqnbs8}
\end{equation}

Considering the case of a 50:50 NPBS, since $|t| = |r|$, using (\ref{eqnbs2}), we get that:
\begin{equation}
|t| = |r| = \frac{1}{\sqrt{2}}.
\label{eqnbs9}
\end{equation}

Further, expressing $|t|$ as $t e^{i \theta}$ and $|r|$ as $r e^{i \phi}$ and using (\ref{eqnbs3}), we get that:
\begin{equation*}
|t| |r| e^{i (\theta - \phi)} + |r| |t| e^{i (\phi - \theta)} = 0
\end{equation*}
\begin{equation*}
\implies 2 |t| |r| \left( \frac{e^{i (\theta - \phi)} + e^{-i (\theta - \phi)}}{2} \right) = 0
\end{equation*}
\begin{equation*}
\implies 2 |t| |r| \cos{(\theta - \phi)} = 0.
\end{equation*}

Using (\ref{eqnbs9}), we get that:
\begin{equation*}
\cos{(\theta - \phi)} = 0.
\end{equation*}

Hence, 
\begin{equation*}
\theta - \phi = \pm \frac{\pi}{2}.
\end{equation*}

Since global phase factors are irrelevant, we can safely set $\theta = 0$ and $\phi = \frac{\pi}{2}$. Equations (\ref{eqnbs7}) and (\ref{eqnbs8}) thus simplify to:
\begin{equation}
\hat{a}^\dag_1 = \frac{1}{\sqrt{2}} (\hat{a}^\dag_3 + i \hat{a}^\dag_4),
\label{eqnbs10}
\end{equation}
and
\begin{equation}
\hat{a}^\dag_2 = \frac{1}{\sqrt{2}} (i \hat{a}^\dag_3 + \hat{a}^\dag_4).
\label{eqnbs11}
\end{equation}

To examine the action of an NPBS on photons received via its 1\textsuperscript{st} and/or 2\textsuperscript{nd} input port(s), we make use of the photon number states (\textit{Fock} states). Let $\ket{\boldsymbol{0}}$ denote the vacuum state and $\ket{\boldsymbol{1}}$ denote the single photon state.

It's evident that when no photons are incident on the NBPS, no photons would emerge from the output ports of the NPBS. Mathematically, it can be said that:
\begin{equation}
\ket{\boldsymbol{0}}_1 \ket{\boldsymbol{0}}_2 \xrightarrow{\text{NPBS}} \ket{\boldsymbol{0}}_3 \ket{\boldsymbol{0}}_4.
\label{eqnbs12}
\end{equation}

Let us consider the case when only 1 photon is incident on a 50:50 NBPS, say, at its 1\textsuperscript{st} input port. Thus, the state at the input of the 50:50 NBPS is $\ket{\boldsymbol{1}}_1 \ket{\boldsymbol{0}}_2 \equiv \hat{a}^\dag_1 \ket{\boldsymbol{0}}_1 \ket{\boldsymbol{0}}_2$. Using (\ref{eqnbs10}) and (\ref{eqnbs12}), we get that:
\begin{align*}
\hat{a}^\dag_1 \ket{\boldsymbol{0}}_1 \ket{\boldsymbol{0}}_2 & \xrightarrow{\text{NPBS}} \frac{1}{\sqrt{2}} (\hat{a}^\dag_3 + i \hat{a}^\dag_4) \ket{\boldsymbol{0}}_3 \ket{\boldsymbol{0}}_4\\
& = \frac{1}{\sqrt{2}} (\ket{\boldsymbol{1}}_3 \ket{\boldsymbol{0}}_4 + i \ket{\boldsymbol{0}}_3 \ket{\boldsymbol{1}}_4).
\end{align*}

Hence, the probabilities of obtaining the above 2 states at the output end of the 50:50 NPBS are given by:
\begin{equation*}
P(\text{Output} = \ket{\boldsymbol{1}}_3 \ket{\boldsymbol{0}}_4) = \left| \frac{1}{\sqrt{2}} \right|^2 = \frac{1}{2},
\end{equation*}
and
\begin{equation*}
P(\text{Output} = \ket{\boldsymbol{0}}_3 \ket{\boldsymbol{1}}_4) = \left| \frac{i}{\sqrt{2}} \right|^2 = \frac{1}{2}.
\end{equation*}
which clearly means that the transmission and reflection of a single photon incident on a 50:50 NPBS are equiprobable.

Now, let us consider the case when 2 \textit{identical} photons are simultaneously incident on the 50:50 NPBS at its 2 different input ports. Thus, the state at the input of the 50:50 NPBS is $\ket{\boldsymbol{1}}_1 \ket{\boldsymbol{1}}_2 \equiv \hat{a}^\dag_1 \hat{a}^\dag_2 \ket{\boldsymbol{0}}_1 \ket{\boldsymbol{0}}_2$. Using (\ref{eqnbs10}),  (\ref{eqnbs11}) and (\ref{eqnbs12}), we get that:
\begin{align}
\hat{a}^\dag_1 \hat{a}^\dag_2 \ket{\boldsymbol{0}}_1 \ket{\boldsymbol{0}}_2 & \xrightarrow{\text{NPBS}} \frac{1}{\sqrt{2}} (\hat{a}^\dag_3 + i \hat{a}^\dag_4) \frac{1}{\sqrt{2}} (i \hat{a}^\dag_3 + \hat{a}^\dag_4) \ket{\boldsymbol{0}}_3 \ket{\boldsymbol{0}}_4 \nonumber\\
& = \frac{1}{2} (i \ket{\boldsymbol{2}}_3 \ket{\boldsymbol{0}}_4 + \cancel{\ket{\boldsymbol{1}}_3 \ket{\boldsymbol{1}}_4} - \cancel{\ket{\boldsymbol{1}}_3 \ket{\boldsymbol{1}}_4} \nonumber\\ & \ \quad \quad + i \ket{\boldsymbol{0}}_3 \ket{\boldsymbol{2}}_4)\label{eqnbs13}\\
& = \frac{i}{\sqrt{2}} (\ket{\boldsymbol{2}}_3 \ket{\boldsymbol{0}}_4 + \ket{\boldsymbol{0}}_3 \ket{\boldsymbol{2}}_4).\label{eqnbs14}
\end{align}

Thus, the 2 photons can be transmitted or reflected \textit{together} with the same probability. It is however, not possible for one of the photons to be reflected and the other one to be transmitted. This is the Hong-Ou-Mandel (HOM) effect~\cite{b24}. 

Note that the 2 terms: $\ket{\boldsymbol{1}}_3 \ket{\boldsymbol{1}}_4$ and $ - \ket{\boldsymbol{1}}_3 \ket{\boldsymbol{1}}_4$ could be cancelled only because the 2 photons were identical. If this had not been the case, say, for instance, when the 2 photons were of 2 different polarizations, then these 2 terms would not have been cancelled and it would have been possible for the 2 input photons to exit through 2 different output ports of the 50:50 NPBS. 

Next, let us consider how the Bell states: $\ket{\phi^\pm}$ and $\ket{\psi^\pm}$ are transformed by a 50:50 NPBS. In order to do this, let us examine how all the 4 possible combinations of orthogonal polarizations of 2 photons (incident simultaneously at the 2 different input ports of a 50:50 NPBS) affect the transformation performed by the 50:50 NPBS. Using (\ref{eqnbs13}) and (\ref{eqnbs14}), we get that (in normalized form):

\begin{equation}
\ket{0}_1 \ket{0}_2 \xrightarrow{\text{NPBS}} \frac{1}{\sqrt{2}} (\ket{0}_3 \ket{0}_3 + \ket{0}_4 \ket{0}_4),
\label{eqnbs15}
\end{equation}
\begin{equation}
\ket{1}_1 \ket{1}_2 \xrightarrow{\text{NPBS}} \frac{1}{\sqrt{2}} (\ket{1}_3 \ket{1}_3 + \ket{1}_4 \ket{1}_4),
\label{eqnbs16}
\end{equation}
\begin{equation}
\ket{0}_1 \ket{1}_2 \xrightarrow{\text{NPBS}} \frac{1}{2} (i \ket{0}_3 \ket{1}_3 + \ket{0}_3 \ket{1}_4 - \ket{1}_3 \ket{0}_4 + i \ket{0}_4 \ket{1}_4),
\label{eqnbs17}
\end{equation}
and
\begin{equation}
\ket{1}_1 \ket{0}_2 \xrightarrow{\text{NPBS}} \frac{1}{2} (i \ket{1}_3 \ket{0}_3 + \ket{1}_3 \ket{0}_4 - \ket{0}_3 \ket{1}_4 + i \ket{1}_4 \ket{0}_4).
\label{eqnbs18}
\end{equation}

Consequently, using (\ref{eqna23}) and (\ref{eqna24}), we get that:
\begin{equation}
\ket{\phi^+} \xrightarrow{\text{NPBS}} \frac{1}{2} (\ket{0}_3 \ket{0}_3 + \ket{0}_4 \ket{0}_4 + \ket{1}_3 \ket{1}_3 + \ket{1}_4 \ket{1}_4),
\label{eqnbs19}
\end{equation}
\begin{equation}
\ket{\phi^-} \xrightarrow{\text{NPBS}} \frac{1}{2} (\ket{0}_3 \ket{0}_3 + \ket{0}_4 \ket{0}_4 - \ket{1}_3 \ket{1}_3 - \ket{1}_4 \ket{1}_4),
\label{eqnbs20}
\end{equation}
\begin{equation}
\ket{\psi^+} \xrightarrow{\text{NPBS}} \frac{1}{\sqrt{2}}(\ket{0}_3 \ket{1}_3 + \ket{0}_4 \ket{1}_4),
\label{eqnbs21}
\end{equation}
and
\begin{equation}
\ket{\psi^-} \xrightarrow{\text{NPBS}} \frac{1}{\sqrt{2}}(\ket{0}_3 \ket{1}_4 - \ket{1}_3 \ket{0}_4).
\label{eqnbs22}
\end{equation}

These results form the basis of Bell State Measurement (BSM). We discuss their implications in detail in Section~\ref{QTimpl} 

All these aforementioned cases have been carefully considered and implemented in ShaNQar's \textbf{NonPolarizingBeamSplitter} class which models the NPBS.

\subsubsection{Polarizing Beam Splitter}
\label{PBS}

Classically, polarizing beam splitters (PBSs) split an incoming beam of light into 2 orthogonal states of polarization by virtue of their birefringence. In the quantum mechanical realm, however, the operation of a PBS becomes probabilistic. This is what is accomplished by the \textbf{PolarizingBeamSplitter} class in ShaNQar.

Let us consider the case that a PBS is manufactured such that it transmits horizontally polarized light and reflects vertically polarized light. Then, the probability of an incoming photon being reflected by the PBS is dependent upon the extent to which it's vertically polarized, or, equivalently, the probability of the photon being in the vertically polarized state $\ket{V}$. What this means is that, for e.g., while photons possessing a quantum state $\ket{\psi} = \ket{V}$ will always be perfectly reflected by such a PBS since their reflection probability is 1, photons possessing a quantum state $\ket{\psi'} = \frac{\ket{H} + \ket{V}}{\sqrt{2}}$ will be reflected half of the time and transmitted half of the time on average by the PBS since their transmission and reflection probabilities are equal. 

In order to take into account the non-ideal behaviour of the PBS, we have included the Extinction Ratio (ER) of the PBS as a user-configurable attribute. The ER is defined as the ratio of the transmission of horizontal polarization to the transmission of vertical polarization and based on its value, some incoming photons can, for instance, get transmitted even though they are vertically polarized.

A PBS proves to be very useful when there is a requirement of detecting and recording the counts of only single photons of a particular quantum state at only 1 particular detector. Such a requirement naturally arises in QKD when photons of different quantum states are to be measured in different bases.

\subsection{Detector}
\label{detector}

The \textbf{Detector} class in ShaNQar models a single photon detector (SPD) which can detect photons and record their total count with precise timing. 

The intrinsic detection efficiency ($\eta_{det, i}$) is the probability of a photon successfully triggering a count upon being incident on the detector. The overall detection efficiency of the SPD, however, depends upon other important factors as well. Firstly, its coupling efficiency with the corresponding source ($\eta^D_{c}$) must be considered so as to account for any possible losses which may lead to an incident photon not being absorbed by the SPD at all. Secondly, the dead time ($\tau_{d}$), i.e., the recovery time for a given detector needs to be taken into account since during this period of time, the detector cannot possibly differentiate between 2 separate photon detection triggers and hence, cannot register the count of another photon incident on it given the fact that it has just absorbed and detected a photon. $\tau_{d}$ is mostly due to the counting electronics of the detection setup. Apart from $\tau_{d}$, the timing jitter ($\tau_{j}$) of the detector must also be taken into account before deciding the time when the next incident photon can possibly be detected.  $\tau_{j}$ captures the standard deviation in the time required for the generation of an electrical pulse by the detector post the absorption of a photon~\cite{b25}. Accordingly, the next instant of time when another incident photon can possibly be detected by the SPD is decided by the time interval $T_{next}$ which follows the instant of time at which the successful detection of a photon has already (but just previously) taken place. This is given by:
\begin{equation}
    T_{next} = \tau_{d} + \tau_{j}r_{norm}, 
\end{equation}
where, $r_{norm} \sim \mathcal{N}(0,\,1)$.

In addition to all this, another important parameter to be necessarily considered is the dark count rate of the detector ($R_{dark}$) which leads to the triggering of false photon counts. Dark counts generally have a thermal origin but may also occur due to material properties such as defects and biasing conditions of the detector. Experimentally, it has been found that dark counts generally originate from a poisson distribution~\cite{b26}. The time interval between dark counts thus follows an exponential distribution. All the dark counts generated during the detector's operation are recorded (stored) and examined while performing the detection analysis of photon(s) (see Sections~\ref{SPDA} and~\ref{BSDA}). 

All the aforementioned characteristic attributes have been carefully considered in our model of the SPD.

\subsection{Node}
\label{node}

The \textbf{Node} class in ShaNQar models a node in a quantum network. In a broad sense, a node is where it all starts (\emph{sender}) and where it all ends (\emph{receiver}). What this basically means, is that, a sender/receiver node essentially consists of all the components that might be required to be operated upon by the sender/receiver in a quantum network. Components assigned to a particular node can thus, directly be controlled from the node in our simulator once all such assignments have taken place. Further, nodes can be connected via classical channels so as to send and/or receive classical information. 

Since experimental quantum networks generally contain more than 1 detectors, detection of the quantum state of photon(s) requires careful consideration of detection timing and dark counts of all the detectors involved. Consequently, our \textbf{Node} class also contains functions for affecting a detection analysis of photon(s). Such a detection analysis proceeds by first determining the cutoffs for the detection events, i.e., the minimum and maximum photon transmission times. Clearly, the \textit{mean} photon transmission time ($\mu_t$) can be estimated by summing up all the \textit{ideal} photon transmission times corresponding to all the components that can be possibly traversed by a photon in the quantum network. We use a user-specified threshold $k$ as a standard deviation ($\sigma^2_t$) multiplier and accordingly, determine the minimum photon transmission time as $\mu_t - k \sigma^2_t$ and the maximum photon transmission time as $\mu_t + k \sigma^2_t$. Evidently, $\sigma^2_t$ is nothing but the sum of the relevant temporal widths of a photon which correspond to all the components that can be possibly traversed by it in the quantum network.

Post the determination of the cutoff times, any 1 of the following 2 types of detection analyses can be employed (as would be possible for a given experimental quantum network):

\subsubsection{Single Photon Detection Analysis}
\label{SPDA}

In this case of only a single photon incident on any given detector (out of a set of detectors), it is evident that \textit{at the most}, only 1 of the detectors will register a \textit{true} photon count. However, at the same time, it is possible for any subset or even the entire set of detectors to register \textit{candidate} false photon triggers (dark counts). Candidate false photon triggers are those false photon triggers that are registered in the time interval bounded by the minimum and maximum photon transmission times. Accordingly, in order to determine which detector's photon count will be considered in the course of an experiment, we first determine the earliest candidate false photon trigger. The following 4 cases can occur:

\begin{itemize}
    \item \textbf{Case A (1 True and Non-Zero Number of Candidate False Photon Triggers)}: If the (earliest) candidate false photon trigger occurs before the true photon trigger, then the detection event corresponding to the detector which registered that (earliest) candidate false photon trigger would be considered. If that is not the case, then the detection event corresponding to the detector which registered the true photon trigger would be considered. 
    \item \textbf{Case B (1 True and Zero Candidate False Photon Triggers)}: Clearly, the detection event corresponding to the detector which registered that true photon trigger would be considered. 
    \item \textbf{Case C (Zero True and Non-Zero Number of Candidate False Photon Triggers)}: Clearly, the detection event corresponding to the detector which registered that (earliest) candidate false photon trigger would be considered. 
    \item \textbf{Case D (Zero True and Zero Candidate False Photon Triggers)}: Evidently, it would be established that no photon has been detected at all. 
\end{itemize}

We use this detection analysis in our experimental QKD simulation (see Section~\ref{QKDimpl}).  

Note that \textit{true} and \textit{false} photon triggers have been separately considered for only the purpose of accurate simulation at the SPD level. Affecting the detection analysis described above ensures that the end detection result is the same as that for an experimentalist who cannot possibly distinguish between \textit{true} and \textit{false} photon triggers (and hence considers both of them as true) at the single photon level.

\subsubsection{Bell State Detection Analysis}
\label{BSDA}

In this case, \textit{at the most} 2 photons can be simultaneously incident on the same detector or 2 different detectors. Here, we begin by determining the 2 earliest \textit{candidate} false photon triggers. The following cases can occur:

\begin{itemize}
    \item \textbf{Case A (2 True and $\mathbf{\geq 2}$ Candidate False Photon Triggers)}: 
    \begin{enumerate}[label=(\alph*)]
        \item If the 2 (earliest) candidate false photon triggers occur before the 2 true photon triggers, then the simultaneous detection event corresponding to the detectors which registered those 2 (earliest) candidate false photon triggers would be considered.
        \item If 1 of the true photon triggers occurs before one of the 2 (earliest) candidate false photon triggers, then the simultaneous detection event corresponding to the detector which registered that earliest true photon trigger and the detector which registered the earliest (out of the $\geq 2$) candidate false photon triggers would be considered.
        \item If the 2 true photon triggers occur before the 2 (earliest) candidate false photon triggers, then the simultaneous detection event corresponding to the detectors which registered those 2 true photon triggers would be considered.
    \end{enumerate} 
    \item \textbf{Case B (2 True and 1 Candidate False Photon Triggers)}:
    \begin{enumerate}[label=(\alph*)]
        \item If the candidate false photon trigger occurs before one of the 2 true photon triggers, then the simultaneous detection event corresponding to the detector which registered that candidate false photon trigger and the detector which registered the earliest (out of the 2) true photon triggers would be considered.
        \item If the 2 true photon triggers occur before the candidate false photon trigger, then the simultaneous detection event corresponding to the detectors which registered those 2 true photon triggers would be considered.
    \end{enumerate}
    \item \textbf{Case C (1 True and $\mathbf{\geq 2}$ Candidate False Photon Triggers)}:
    \begin{enumerate}[label=(\alph*)]
        \item If the true photon trigger occurs before one of the 2 (earliest) candidate false photon triggers, then the simultaneous detection event corresponding to the detector which registered that true photon trigger and the detector which registered the earliest (out of the $\geq 2$) candidate false photon triggers would be considered.
        \item If the 2 (earliest) candidate false photon triggers occur before the true photon trigger, then the simultaneous detection event corresponding to the detectors which registered those 2 (earliest) candidate false photon triggers would be considered.
    \end{enumerate} 
    \item \textbf{Case D (1 True and 1 Candidate False Photon Triggers)}: Clearly, the simultaneous detection event corresponding to the detector which registered that true photon trigger and the detector which registered that candidate false photon triggers would be considered.
    \item \textbf{Case E (Zero True and $\mathbf{\geq 2}$ Candidate False Photon Triggers)}: Clearly, the simultaneous detection event corresponding to the detectors which registered those 2 (earliest) candidate false photon triggers would be considered.
    \item \textbf{Case F (1 or Zero (True or False) Photon Triggers)}: Evidently, it would be established that no bell state (entangled photon pair) has been detected at all. 
\end{itemize}

We use this detection analysis in our experimental quantum teleportation simulation (see Section~\ref{QTimpl}).

\begin{figure}[htbp]
\begin{center}
\includegraphics[scale = 0.70]{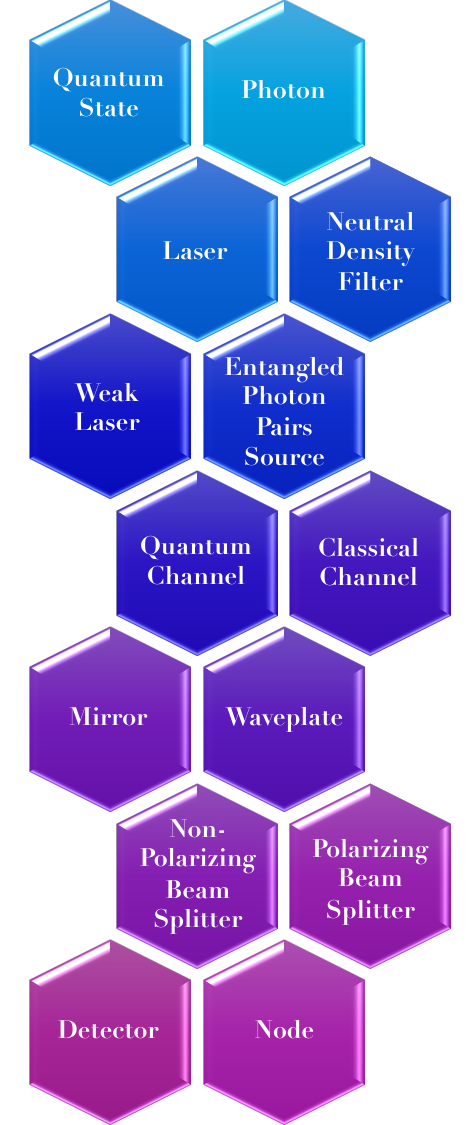}
\end{center}
\caption{Overview of ShaNQar's main components}
\label{main_overview}
\end{figure}

Having explained the key components of ShaNQar, we now give a diagrammatic overview of the same in Fig.~\ref{main_overview}. The honeycomb-like structure in Fig.~\ref{main_overview} presents all the components of ShaNQar in an ordered fashion starting from foundational blocks at the top (\textbf{QuantumState} and \textbf{Photon}), progressively building up on them with subsequent components, and finally ending at a cohesion of all of them in the form of the \textbf{Node} class. 

In the subsequent sections, we simulate real-life experimental setups pertaining to quantum key distribution (Section~\ref{QKD}) and quantum teleportation (Section~\ref{QT}).

\section{Quantum Key Distribution}
\label{QKD}

\subsection{Theory}
\label{QKDth}
Quantum Key Distribution (QKD) is a more secure alternative to the \textit{vulnerable} classical methods of key distribution such as the DH~\cite{b27} and RSA~\cite{b28} protocols. It is a technique for the generation of an encryption/decryption key which is shared between a sender(Alice) and a receiver(Bob). It utilizes the state destruction property of quantum-mechanical measurements to its advantage. More formally, by ensuring that only non-orthogonal states are transmitted from Alice to Bob, it allows us to obtain a sure-shot guarantee that an eavesdropper Eve cannot clone the quantum states of the photons being sent by Alice to Bob by virtue of the veracity of the no-cloning theorem. Further, as mentioned earlier, Eve cannot gain any amount of information from the non-orthogonal states without perturbing them. Such measurements performed by Eve would thus inevitably introduce \textit{detectable} modifications in the key. Consequently, the QKD protocol can be aborted midway by Alice and Bob when the modifications in the key are greater than a predefined threshold and the protocol can then be reinitiated. Evidently, QKD's security is guaranteed by quantum-mechanical principles which revolve around the fundamental laws of physics~\cite{b29}!

Inspired by the work of Wiesner in the 1970s~\cite{b30}, \textbf{B}ennett and \textbf{B}rassard proposed the very first seminal protocol for QKD in 19\textbf{84}~\cite{b29}. This protocol hence came to be known as the \textbf{BB84} protocol. We now enlist the steps involved in this protocol~\cite{b1, b29} and describe them in great detail.

\subsubsection{BB84 Protocol}
\label{BB84}

\begin{enumerate}
    \item Alice chooses 2 random \emph{classical} bit strings of length $N_{tr}$, say, $p$ and $q$ and then encodes each bit of $p$ in the basis indicated by the corresponding bit of $q$. Thus, $q$ represents a random string of polarization bases: a 0 represents the Rectilinear (Horizontal/Vertical) or $Z$ basis while a 1 represents the Diagonal (Diagonal/Anti-Diagonal) or $X$ basis. Mathematically, the scheme for encoding the $i\textsuperscript{th}$ bit of $p$ ($= p_i$) in the corresponding bit of $q$ ($= q_i$) and hence, obtaining the corresponding polarization (quantum) state of the photon to be transmitted to Bob ($ = \ket{\psi_{p_i q_i}}$) works as follows:
    \begin{enumerate}
        \item When $p_i = 0$ and $q_i = 0$, $\ket{\psi_{00}} = \ket{0}$
        \item When $p_i = 1$ and $q_i = 0$, $\ket{\psi_{10}} = \ket{1}$
        \item When $p_i = 0$ and $q_i = 1$, $\ket{\psi_{01}} = \ket{+}$
        \item When $p_i = 1$ and $q_i = 1$, $\ket{\psi_{11}} = \ket{-}$
    \end{enumerate}
    Consequently, she prepares the $i\textsuperscript{th}$ photon in the quantum state $\ket{\psi_{p_i q_i}}$ and then sends it across (via the physical QKD system) to Bob. Clearly, all these 4 different types of states are mutually non-orthogonal and therefore, distinguishability between all of them is impeded by the laws of quantum  mechanics.
    \item Corresponding to each quantum state $\ket{\psi_{p_i q_i}}$ sent by Alice, Bob receives the state $\mathcal{E}(\ket{\psi_{p_i q_i}}\!\bra{\psi_{p_i q_i}})$, where, $\mathcal{E}$ refers to the quantum operation encapsulating the effects of both (any possible) eavesdropping and (the ubiquitous) noise. It is important to note that in a practical scenario, in certain cases, Bob may not receive a photon at all at his (detection) side due to various losses (majorly, the limited coupling and transmission efficiencies) involved during the transmission of photons through the QKD system. Nevertheless, the photons which are finally received by Bob are measured by him in a polarization basis ($X$ or $Z$) chosen at random. Together, these bases form a random bit string $q'$. Moreover, the measured quantum states in the corresponding basis form a bit string $p'$ when considered together. Effectively, $p'$ is Bob's \textbf{\emph{raw} key string}. Evidently, $p' \neq p$.
    \item With the help of a public (classical) channel, Alice then communicates the value of $q$ to Bob. At his end, Bob checks the indices of the bit strings $q$ and $q'$ so as to figure out a list of indices $q_{idx, same}$ such that $\forall\,k \in q_{idx, same} ,\,q_k = q'_k$. He then sends across $q_{idx, same}$ to Alice via the same public channel. Subsequently, Alice and Bob retain only those bits in $p$ and $p'$ whose corresponding indices are elements of the list $q_{idx, same}$ (the rest of the bits are hence, discarded by both of them). Let these bits together form bit strings $\widetilde{p}$ and $\widetilde{p'}$ respectively of the same length $N_{sift}$. Thus, $\widetilde{p}$ and $\widetilde{p'}$ are the \textbf{\emph{sifted} key strings}. Note however that $\widetilde{p} \neq \widetilde{p'}$ due to the effects of noise and any eavesdropping by Eve.
    \item Alice then randomly chooses a half of the bits in $\widetilde{p}$ such that their indices (together) form a list $c$ of length $\left \lfloor \frac{N_{sift}}{2} \right \rfloor$. $c$ is essentially the check list which serves as a check on any unacceptable level of noise and eavesdropping effects. Alice and Bob then share with the (respective) other all those elements of $\widetilde{p}$ and $\widetilde{p'}$ (respectively) whose indices lie in $c$. By comparing the values of these corresponding bit elements and figuring out the number of bits $e$ possessing different values, Alice and Bob are able to estimate the value of the Quantum Bit Error Rate (QBER). If the QBER is greater than a pre-determined threshold, they abort the QKD protocol. Otherwise, from here onwards, they only consider those bits of $\widetilde{p}$ and $\widetilde{p'}$ (respectively) whose indices do not lie in $c$. Let these bits together form bit strings $s$ and $s'$ respectively of the same length $\left \lceil \frac{N_{sift}}{2} \right \rceil$. 
    \item Evidently, even $s$ and $s'$ wouldn't be completely equal to one another. In order to correct errors in $s'$ and hence, obtain a string $\widetilde{s'} \approx s$, Alice and Bob perform information reconciliation. Effectively, $\widetilde{s'}$ is Bob's \textbf{\emph{reconciled} key string}.
    \item The information reconciliation stage does involve the exchange of some information regarding the key shared by Alice and Bob over a public channel which may give an eavesdropper additional information to construct the same key. Consequently, Alice and Bob use a process known as privacy amplification so as to condense $s$ and $\widetilde{s'}$ to smaller keys $x$ and $x'$, whose correlation with the leaked information is very low. $x$ and $x'$ are thus, Alice's and Bob's \textbf{\emph{secret} keys}!  
\end{enumerate}
ShaNQar implements information reconciliation by using an optimized version of the Cascade Protocol detailed by Choudhary and Wasan in~\cite{b31}. We will elaborate on ShaNQar's implementation of Privacy Amplification in the next subsection.

\subsubsection{Privacy Amplification}
\label{PAmp}

As noted in the previous subsection, an eavesdropper Eve can gain information via passive eavesdropping in the information reconciliation stage. In order to compensate for this loss of information of say, $f$ bits of the key, we perform privacy amplification so as to reduce/shrink Alice's and Bob's key strings by $f$ bits. This is accomplished with the help of universal hashing so as to minimize the collision probability associated with different initial \textit{long} key strings corresponding to the same final \textit{compressed} key string. The collision probability should hence be minimized to as great an extent as possible so as to ensure the secrecy of the final key~\cite{b32}. 

Toeplitz matrices are the most preferred kind of universal\textsubscript{2} hash (The subscript of 2 indicates that the function maps one binary string to another, i.e., from $\{0,1\}^i \rightarrow \{0,1\}^f$) functions since they can be specified by only a small number of randomly chosen bits. A Toeplitz matrix of size $f \times i$ (where, $f$ = final \emph{compressed} key length and $i$ = initial \emph{long} key length) is given by~\cite{b33}:
\begin{equation}
    T = \begin{bmatrix}
        r_{0} & r_{-1} & r_{-2} & \cdots & r_{-i + 1} \\
        r_{1} & r_{0} & r_{-1} & \ddots &  \\
        r_{2} & r_{1} & \ddots & & \vdots \\
        \vdots & \ddots &  &  &  \\
        r_{f - 1} &  & \hdots &  & r_{f - i} \\
    \end{bmatrix}.
    \label{eqn_toeplitz}
\end{equation}

From the above equation, it is clear that Toeplitz matrices are essentially  constant diagonal matrices, where, $T_{i,j} = r_{i - j}$, and $r$ is a vector of only $f + i - 1$ random bit values in total. Alice and Bob accomplish hashing by multiplying their key strings by $T$ so as to obtain their respective secret keys (hash values). 

\subsection{Experimental Simulation}
\label{QKDsim}

\subsubsection{Implementation Details}
\label{QKDimpl}

\begin{figure*}[htbp]
\begin{center}
\includegraphics[scale = 0.50]{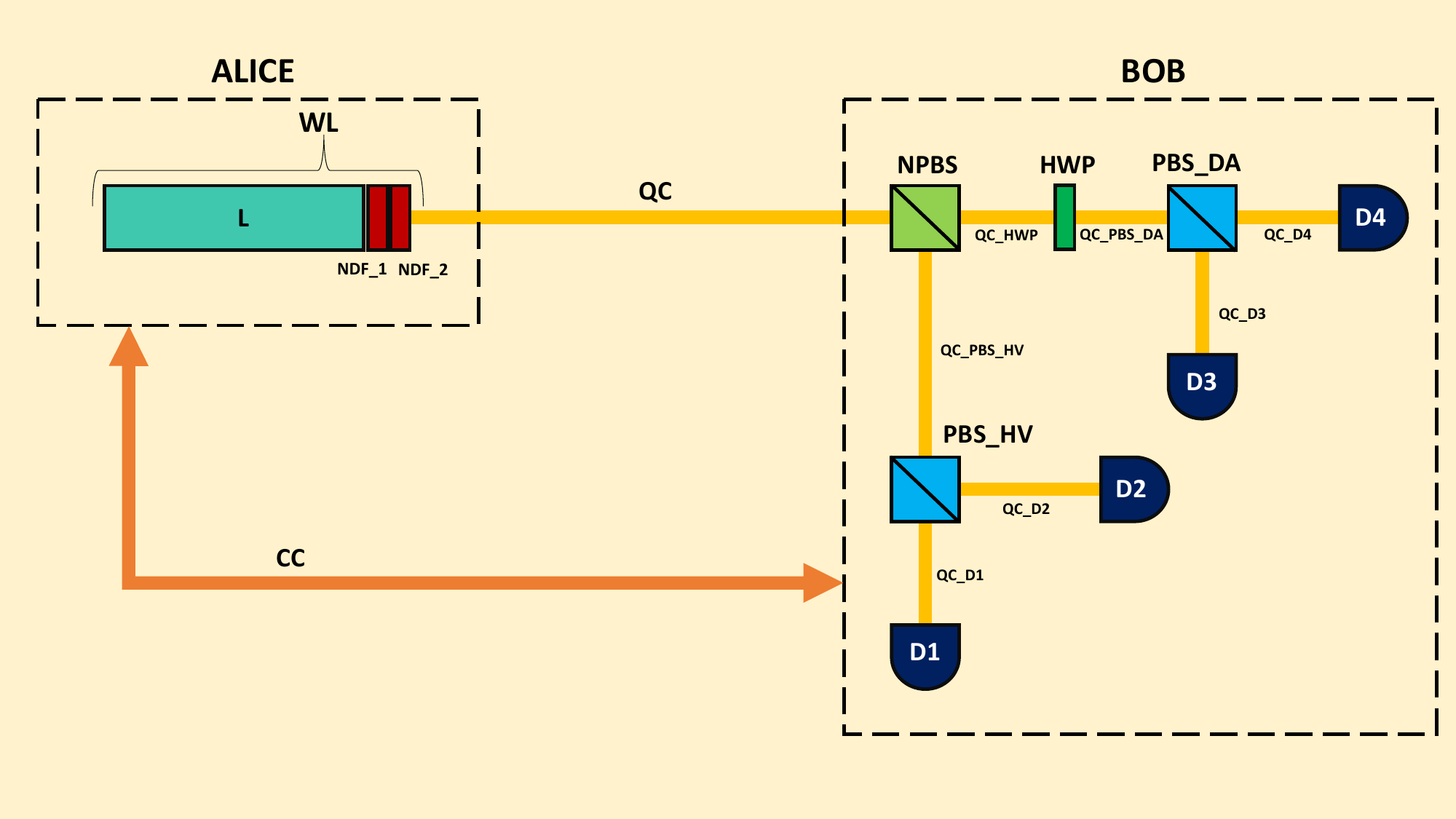}
\end{center}
\caption{Experimental setup for QKD simulation (The setup is not drawn to scale and is rather enlarged in certain areas for better clarity)}
\label{figQKD}
\end{figure*}

The experimental setup, i.e., the all-fiber physical system for QKD that has been simulated by us using ShaNQar is shown in detail in Fig.~\ref{figQKD}. The values of the parameters of the components in Fig.~\ref{figQKD} are given in Table~\ref{table1a} and have been established from the literature~\cite{b34, b1, qkd1}.

\begingroup
\squeezetable
\begin{table}
\centering
\begin{tabular}{|c|l|}
\hline
\textbf{Components}&\textbf{Parameters} \\
\hline
\textbf{WL}           & \begin{tabular}[c]{@{}l@{}}PRR = $80$ MHz\\ $\lambda = 1550$ nm\\ $\Delta \lambda = 0.01$ nm\\ $W^L_{temp} = 100$ fs\\ $\mu_{p} = 10^5$ photons/s\\ NL = $0.01$\\ $\gamma_{amp\_damp} = 0.3$\\ $\lambda_{phase\_damp} = 0.45$\\ $OD_{min} = 0$\\ $OD_{max} = 4$\\ $N_{NDF,max} = 5$\end{tabular} \\ \hline
\textbf{QC}                                 & \begin{tabular}[c]{@{}l@{}}$L = 1$ km\\ $\eta^{QC}_c = 0.85$\\ $\alpha = 0.2$ dB/km\\ $n_{core} = 1.47$\\ $F_{pol} = 0.90$\\ $D_{chr} = 17$ ps/nm-km\\ $p = 0.3$\end{tabular}                                        \\ \hline
\textbf{CC}                                 & \begin{tabular}[c]{@{}l@{}}$L= 1$ km\\ $n_{core} = 1.47$\end{tabular}                                                                                     \\ \hline
\textbf{NPBS}                               & $R = 0.50$                                                                                                                                       \\ \hline
\textbf{PBS\_HV and PBS\_DA}                & ER = $1000$                                                                                                                                       \\ \hline
\textbf{QC\_PBS\_HV}                        & \begin{tabular}[c]{@{}l@{}}$L = 0.50$ m\\ $\eta^{QC}_c = 0.85$\\ $\alpha = 0.2$ dB/km\\ $n_{core} = 1.47$\\ $F_{pol} = 0.90$\\ $D_{chr} = 17$ ps/nm-km\\ $p = 0.3$\end{tabular}                                                           \\ \hline
\textbf{QC\_HWP and QC\_PBS\_DA}            & \begin{tabular}[c]{@{}l@{}}$L= 0.25$ m\\ $\eta^{QC}_c = 0.85$\\ $\alpha = 0.2$ dB/km\\ $n_{core} = 1.47$\\ $F_{pol} = 0.90$\\ $D_{chr}= 17$ ps/nm-km\\ $p = 0.3$\end{tabular}                                                           \\ \hline
\textbf{HWP}                                & \begin{tabular}[c]{@{}l@{}}$\alpha = 180^{\circ}$\\ $\theta = 22.5^{\circ}$\end{tabular}                                                                                   \\ \hline
\textbf{QC\_D1, QC\_D2, QC\_D3, and QC\_D4} & \begin{tabular}[c]{@{}l@{}}$L= 0.15$ m\\ $\eta^{QC}_c = 0.85$\\ $\alpha = 0.2$ dB/km\\ $n_{core} = 1.47$\\ $F_{pol} = 0.90$\\ $D_{chr}= 17$ ps/nm-km\\ $p = 0.3$\end{tabular}                                                           \\ \hline
\textbf{D1, D2, D3, and D4}                  & \begin{tabular}[c]{@{}l@{}}$\eta^{D}_c = 0.90$\\ $\tau_{d} = 10$ ns\\ $\eta_{det, i} = 0.90$\\ $R_{dark} = 100$ Hz\\ $\tau_{j} = 55$ ps\end{tabular}                    \\ \hline
\end{tabular}
\caption{Parameters corresponding to the components used in the simulation of the experimental Setup for QKD shown in Fig.~\ref{figQKD}}
\label{table1a}
\end{table}
\endgroup

Note that the notations used for various parameters corresponding to the different components in Fig.~\ref{figQKD} are consistent with the ones mentioned earlier in Section~\ref{hardware_stack}. This setup has been inspired by a similar setup used by Jain \textit{et al}~\cite{b34}. Apart from differences in the values of the parameters corresponding to various components, the major difference between our setup and the one used by them is that while our setup is all-fiber (and hence, consists of single-mode fibers as quantum channels), they have used free-space quantum channels. While free-space quantum channels might deliver a reasonable level of performance over small distances of say, only a few meters (as was the case for them), evidently, absorption, scattering, and other sources of noise in free-space channels can significantly deteriorate the signal quality. This can result in a significant loss of photons and/or corruption of the quantum states of a large number of transmitted photons, when large distances separate the receiver from the sender. For long transmission distances (of the order of km, as is the case for our experimental setup), it is thus, practical to go for a fiber based QKD system which overcomes the aforementioned limitations of free-space channels to a great extent. We now begin to describe how our system performs QKD.

Single photon pulses emitted by the weak laser \textbf{WL} (here, a laser \textbf{L} along with 2 ND filters (\textbf{NDF\_1} and \textbf{NDF\_2})) at Alice's (the sender's) end of the QKD system are encoded in either the rectilinear or the diagonal basis based on 2 separate random classical bit strings generated by Alice. The encoding scheme used by Alice has already been discussed in detail in Section~\ref{BB84}.

These single photons travel through the long quantum channel \textbf{QC} to possibly finally reach Bob's (the receiver's) end of the QKD system where they are directly incident on the non-polarizing beam splitter \textbf{NPBS}. The 50:50 \textbf{NPBS} effectively performs the function of random yet equiprobable basis selection (for the purpose of measuring the quantum state of the transmitted photons) by either reflecting an incoming photon over to the quantum channel \textbf{QC\_PBS\_HV} or transmitting it to the quantum channel \textbf{QC\_HWP}. The 2 segments starting from the 2 output ports of the \textbf{NPBS} essentially perform the function of measuring the quantum state of the incoming photon in the basis decided by the \textbf{NPBS}. Let us examine all this with the help of concrete examples:
\begin{enumerate}
    \item If an incoming photon with a quantum state $\ket{\psi} \equiv \ket{0}$ or $\ket{\psi} \equiv \ket{1}$ is reflected by the \textbf{NPBS} to the quantum channel \textbf{QC\_PBS\_HV}, it will then be effectively measured in the correct basis by the polarizing beam splitter \textbf{PBS\_HV} positioned at the receiver end of \textbf{QC\_PBS\_HV} (provided that its quantum state is not significantly corrupted by noise or any other eavesdropping effects). Essentially, this is because \textbf{PBS\_HV} will generally transmit a horizontally polarized photon and reflect a vertically polarized photon. This means that a photon with the quantum state $\ket{\psi} \equiv \ket{0}$ will be transmitted to the detector \textbf{D1} via the quantum channel \textbf{QC\_D1} while on the other hand, a photon with the quantum state $\ket{\psi} \equiv \ket{1}$ will be reflected to the detector \textbf{D2} via the quantum channel \textbf{QC\_D2}. Hence, photon counts registered by \textbf{D1} necessarily correspond to the detection of photons with $\ket{\psi} \equiv \ket{0}$ while those registered by \textbf{D2} necessarily correspond to the detection of photons with $\ket{\psi} \equiv \ket{1}$.
    \item If an incoming photon with a quantum state $\ket{\psi} \equiv \ket{+}$ or $\ket{\psi} \equiv \ket{-}$ is reflected by the \textbf{NPBS} to the quantum channel \textbf{QC\_HWP}, it will then be made to pass through the half-waveplate \textbf{HWP} inclined at $\theta = \frac{\pi}{8}$. As noted earlier in Section~\ref{waveplate}, (upto an irrelevant global phase factor) such a half-wave plate acts as a Hadamard ($H$) gate. Consequently, on passing through \textbf{HWP}, the quantum state $\ket{+}$ will be changed to $\ket{0}$ while a quantum state of $\ket{-}$ will be changed to $\ket{1}$. The photon in this new quantum state of either $\ket{0}$ or $\ket{1}$ will then be transmitted over to the polarizing beam splitter \textbf{PBS\_DA} via the quantum channel \textbf{QC\_PBS\_DA}. Again, provided that its quantum state is not significantly corrupted by noise or any other eavesdropping effects, such an incoming photon will be effectively measured in the correct basis by \textbf{PBS\_DA}. Essentially, this is because the orientation of \textbf{PBS\_DA} is such that it will generally reflect a horizontally polarized photon and transmit a vertically polarized photon. This means that a photon with the new quantum state $\ket{\psi'} \equiv \ket{0}$ will be reflected to the detector \textbf{D3} via the quantum channel \textbf{QC\_D3} while on the other hand, a photon with the new quantum state $\ket{\psi'} \equiv \ket{1}$ will be transmitted to the detector \textbf{D4} via the quantum channel \textbf{QC\_D4}. Hence, photon counts registered by \textbf{D3} necessarily correspond to the detection of photons with $\ket{\psi} \equiv \ket{+}$ while those registered by \textbf{D4} necessarily correspond to the detection of photons with $\ket{\psi} \equiv \ket{-}$.
    \item If an incoming photon with a quantum state $\ket{\psi} \equiv \ket{+}$ or $\ket{\psi} \equiv \ket{-}$ is reflected by the \textbf{NPBS} to the quantum channel \textbf{QC\_PBS\_HV}, it will then be effectively measured in the incorrect basis by the polarizing beam splitter \textbf{PBS\_HV} positioned at the receiver end of \textbf{QC\_PBS\_HV}. Further, photons in the quantum states $\ket{\pm} = \frac{1}{\sqrt{2}} (\ket{0} \pm \ket{1})$ have a 50\% probability of being reflected by \textbf{PBS\_HV} to the detector \textbf{D2} (via the quantum channel \textbf{QC\_D2}) and a 50\% probability of being transmitted by \textbf{PBS\_HV} to the detector \textbf{D1} (via the quantum channel \textbf{QC\_D1}). Thus, the measured quantum state coefficients will also turn out to be incorrect $\approx$ 50\% of the times. 
    \item If an incoming photon with a quantum state $\ket{\psi} \equiv \ket{0}$ or $\ket{\psi} \equiv \ket{1}$ is reflected by the \textbf{NPBS} to the quantum channel \textbf{QC\_HWP}, it will then be effectively measured in the incorrect basis by the polarizing beam splitter \textbf{PBS\_DA}. When such photons are made to pass through \textbf{HWP}, their quantum states will be effectively changed to $\ket{+}$ and $\ket{-}$ respectively. Again, it must be noted that photons in the quantum states $\ket{\pm} = \frac{1}{\sqrt{2}} (\ket{0} \pm \ket{1})$ have a 50\% probability of being transmitted by \textbf{PBS\_DA} to the detector \textbf{D4} (via the quantum channel \textbf{QC\_D4}) and a 50\% probability of being reflected by \textbf{PBS\_DA} to the detector \textbf{D3} (via the quantum channel \textbf{QC\_D3}). Thus, the measured quantum state coefficients will also turn out to be incorrect $\approx$ 50\% of the times.
\end{enumerate}

Alice and Bob then follow steps 3, 4, 5, and 6 of the BB84 protocol given in Section~\ref{BB84} so as to possibly obtain their final secret key strings. All these steps involve the usage of the classical channel \textbf{CC}. As mentioned earlier, information reconciliation is performed via an optimized version of the Cascade protocol and Algorithm 3 of~\cite{b31} is implemented. We note that 2 - 3\% QBER values are generally, the minimum average values of QBER achieved by the highly robust and efficient physical QKD systems of today. Accordingly, the error probability $p$ in Cascade is chosen to be 2\% higher than the estimate of QBER obtained from Step 4 of the BB84 protocol so as to compensate for any loss (in the form of no/negligible amount of error correction being performed by Cascade) incurred due to an incorrect estimation of QBER since the random set of bits chosen by Alice for estimation of the QBER can very well turn out to have zero or a very low amount of error while the other half might then have all the errors or contain a very high amount of error. Privacy Amplification is finally performed using Toeplitz matrices (as outlined in Section~\ref{PAmp}) wherein Alice chooses the random vector $r$ and sends it over to Bob via \textbf{CC}. Additionally, we set the value of $f$ to be half the reconciled key string length so as to consider a possible worst case estimate (when Eve knows some information pertaining to almost half of the bits contained in the key). 

\subsubsection{Results}
\label{QKDres}
\begin{figure}[htbp]
\begin{center}
\includegraphics[scale = 0.55]{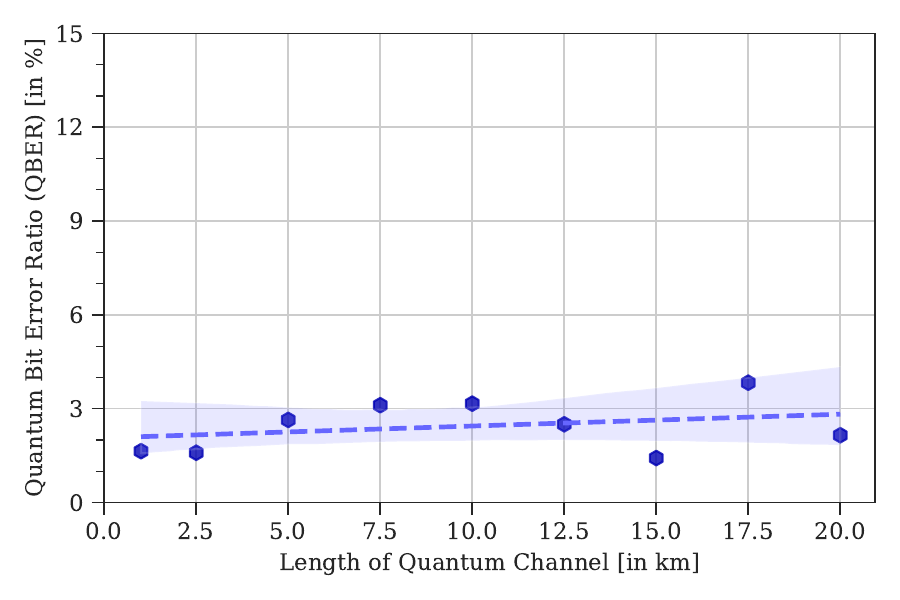}
\end{center}
\caption{QBER v/s QC length for the polarization-encoding based BB84 protocol simulated using the setup illustrated in Fig.~\ref{figQKD}}
\label{QBER}
\end{figure}

\begin{figure}[htbp]
\begin{center}
\includegraphics[scale = 0.55]{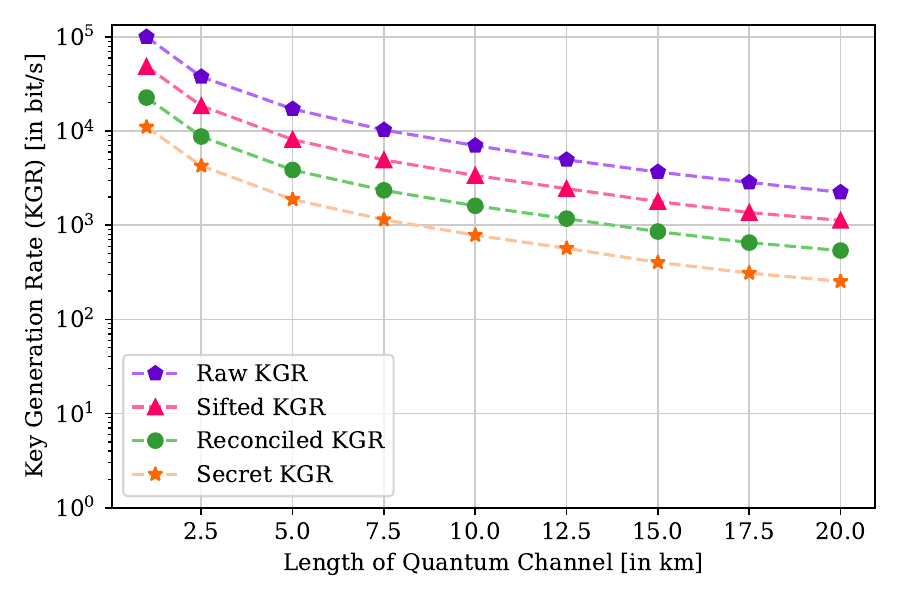}
\end{center}
\caption{KGRs v/s QC length for the polarization-encoding based BB84 protocol simulated using the setup illustrated in Fig.~\ref{figQKD}}
\label{KGR}
\end{figure}

In line with the details mentioned in Section~\ref{QKDimpl}, we simulated QKD using the experimental setup shown in Fig.~\ref{figQKD} while varying the length of the internode quantum channel \textbf{QC} from 0 to 20 km in steps of 2.5 km with 1 km as the starting point (a 0 km \textbf{QC} obviates the need for QKD). We averaged out the results for each \textbf{QC} length over 10 runs. The weak laser \textbf{WL} transmitted 0.2179 photons on average. Accordingly, in an attempt to successfully transmit about 128 (raw key) photons to Bob over a \textbf{QC} of length 1 km (this number would gradually decrease with an increase in the \textbf{QC}'s length) before distillation into the final secret key, Alice fired $\approx$ 10x more single photon pulses, i.e., $2.2\left(\frac{100}{\mu_a}\right)$ pulses in total, where, $\mu_a$ denotes the mean number of photons emitted by \textbf{WL} in a given run. The results of all our experiments are shown in Fig.~\ref{QBER} and Fig.~\ref{KGR}.

From Fig.~\ref{QBER}, we can clearly see that the QBER shows a trend of a slight gradual increase from 1 to 20 km. This is expected given the decrease in the \textbf{QC}'s transmittance $T$ with the increase in its length $L$ (see (\ref{eqna33})). At the same time, we observe that the QBER remains in the range of 1 - 4 \%. 

Fig.~\ref{KGR} reveals that the Key Generation Rates (KGRs), i.e., the number of key bits generated per second, exponentially decrease with the increase in the \textbf{QC}'s length. As expected, Raw KGR $>$ Sifted KGR $>$ Reconciled KGR $>$ Secret KGR (see Section~\ref{QKDth} for interpretations of these terms). This is evident given the distillation at every stage of QKD and the increase in the classical communication overhead which results in a progressive decrease in the number of bits and increase in key generation time all the way from the raw to the secret key strings (respectively). 

All these observations are in line with previous QKD experiments~\cite{qkd1, qkd2}. They also serve as a validation of the fact that ShaNQar is able to accurately account for errors in various components resulting from finite efficiencies, transmittances, polarization extinction ratios, detector dark counts, different types of noise, etc. 

\section{Quantum Teleportation}
\label{QT}

\subsection{Theory}
\label{QTth}

Quantum Teleportation (QT) is a technique for `teleporting' a quantum state, i.e., physically transferring a target quantum state (= $\ket{\psi_T}$) from a sender (Alice) to a receiver (Bob) despite the absence of a quantum channel linking Alice to Bob. QT is made possible with the help of an entangled (Bell) state shared between Alice and Bob and involves some amount of classical communication which takes place from Alice to Bob. Alice interacts $\ket{\psi_T}$ with her half of the Bell state and then performs a joint measurement of the 2 qubits. Although this joint measurement destroys the target qubit in her possession, it allows Bob to recover the original state $\ket{\psi_T}$ at his end after he performs some operations on his half of the Bell state depending upon the measurement's results~\cite{b1, b35}. 

\subsection{Experimental Simulation}
\label{QTsim}

\subsubsection{Implementation Details}
\label{QTimpl}

\begin{figure*}[htbp]
\begin{center}
\includegraphics[scale = 0.50]{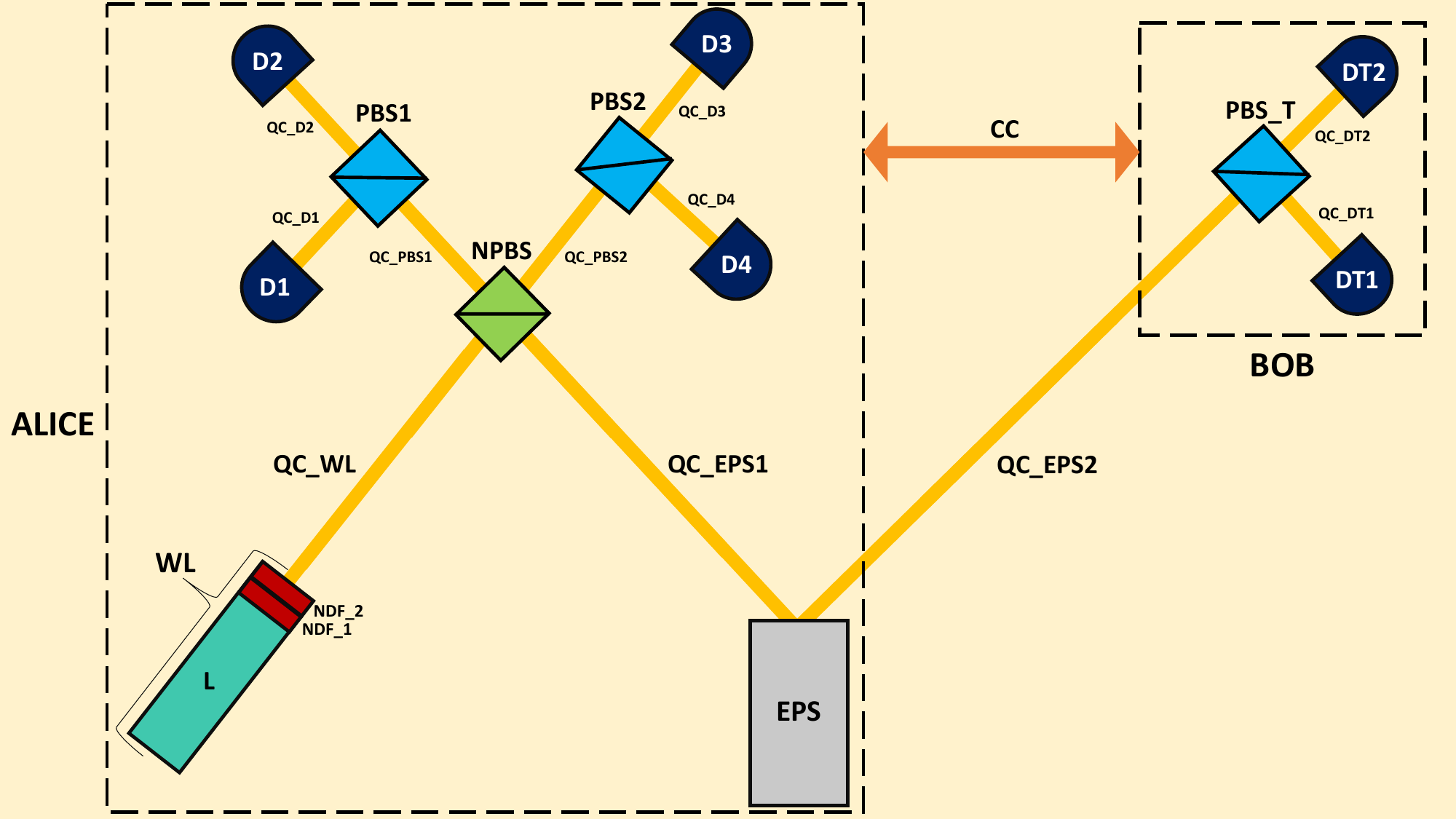}
\end{center}
\caption{Experimental setup for QT simulation (The setup is not drawn to scale and is rather enlarged in certain areas for better clarity)}
\label{figQT}
\end{figure*}

The experimental setup, i.e., the all-fiber physical system for QT that has been simulated by us using ShaNQar is shown in detail in Fig.~\ref{figQT}. The values of the parameters of the components in Fig.~\ref{figQT} are given in Table~\ref{tableQT} and have been established from the literature~\cite{b35, b1, b20}.

\begingroup
\squeezetable
\begin{table}
\centering
\begin{tabular}{|c|l|}
\hline
\textbf{Components}&\textbf{Parameters} \\
\hline
\textbf{EPS}           & \begin{tabular}[c]{@{}l@{}}PRR = $76$ MHz\\ $\lambda = 775$ nm\\ $\Delta \lambda = 0.01$ nm\\ $W^L_{temp} = 200$ fs\\ $\mu_{p} = 10^6$ photons/s\\ NL = $0.01$\\ $\gamma_{amp\_damp} = 0.3$\\ $\lambda_{phase\_damp} = 0.45$\\ SPDC Type = $2$\\ $\eta_{SPDC} = 10^{-6}$\end{tabular} \\ \hline
\textbf{WL}           & \begin{tabular}[c]{@{}l@{}}PRR = $76$ MHz\\ $\lambda = 1550$ nm\\ $\Delta \lambda = 0.01$ nm\\ $W^L_{temp} = 200$ fs\\ $\mu_{p} = 10^5$ photons/s\\ NL = $0.01$\\ $\gamma_{amp\_damp} = 0.3$\\ $\lambda_{phase\_damp} = 0.45$\\ $OD_{min} = 0$\\ $OD_{max} = 4$\\ $N_{NDF,max} = 5$\end{tabular} \\ \hline
\textbf{QC\_EPS1 and QC\_WL}            & \begin{tabular}[c]{@{}l@{}}$L = 100$ m\\ $\eta^{QC}_c = 0.95$\\ $\alpha = 0.2$ dB/km\\ $n_{core} = 1.47$\\ $F_{pol} = 0.90$\\ $D_{chr}= 17$ ps/nm-km\\ $p = 0.3$\end{tabular}                                                           \\ \hline
\textbf{QC\_EPS2}                                 & \begin{tabular}[c]{@{}l@{}}$L = 1$ km\\ $\eta^{QC}_c = 0.95$\\ $\alpha = 0.2$ dB/km\\ $n_{core} = 1.47$\\ $F_{pol} = 0.90$\\ $D_{chr} = 17$ ps/nm-km\\ $p = 0.3$\end{tabular}                                        \\ \hline
\textbf{CC}                                 & \begin{tabular}[c]{@{}l@{}}$L= 1$ km\\ $n_{core} = 1.47$\end{tabular}                                                                                     \\ \hline
\textbf{NPBS}                               & $R = 0.50$                                                                                                                                       \\ \hline
\textbf{PBS1, PBS2 and PBS\_T}                & ER = $1000$                                                                                                                                       \\ \hline
\textbf{QC\_PBS1 and QC\_PBS2}            & \begin{tabular}[c]{@{}l@{}}$L= 0.5$ m\\ $\eta^{QC}_c = 0.95$\\ $\alpha = 0.2$ dB/km\\ $n_{core} = 1.47$\\ $F_{pol} = 0.90$\\ $D_{chr}= 17$ ps/nm-km\\ $p = 0.3$\end{tabular}                                                           \\ \hline
\textbf{\shortstack{QC\_D1, QC\_D2, QC\_D3, QC\_D4,\\QC\_DT1 and QC\_DT2}} & \begin{tabular}[c]{@{}l@{}}$L= 0.15$ m\\ $\eta^{QC}_c = 0.95$\\ $\alpha = 0.2$ dB/km\\ $n_{core} = 1.47$\\ $F_{pol} = 0.90$\\ $D_{chr}= 17$ ps/nm-km\\ $p = 0.3$\end{tabular}                                                           \\ \hline
\textbf{D1, D2, D3, D4, DT1 and DT2}                  & \begin{tabular}[c]{@{}l@{}}$\eta^{D}_c = 0.95$\\ $\tau_{d} = 10$ ns\\ $\eta_{det, i} = 0.90$\\ $R_{dark} = 1000$ Hz\\ $\tau_{j} = 55$ ps\end{tabular}                    \\ \hline
\end{tabular}
\caption{Parameters corresponding to the components used in the simulation of the experimental setup for QT shown in Fig.~\ref{figQT}}
\label{tableQT}
\end{table}
\endgroup

Note that the notations used for various parameters corresponding to the different components in Fig.~\ref{figQT} are consistent with the ones mentioned earlier in Section~\ref{hardware_stack}. This setup has been inspired by a similar setup used by Bouwmeester \textit{et al}~\cite{b35}. 

Throughout this simulation, wherever necessary, we use the \textit{adaptive} environment setting. Since 2 photons (in the maximally entangled (Bell) state) are simultaneously generated and can simultaneously propagate through the same/different components of the quantum teleportation network, timing synchronization must take place for the 2 photons individually and a global environment will not suffice in itself. Accordingly, timing information is stored in each photon's environment separately and wherever necessary, a component's environment adapts to the photon's environment propagating through it, i.e., temporarily, its environment becomes the photon's environment. The global environment is synchronized with the local photon environments (as is necessary) at the end of each run of the simulation. 

First of all, an entangled photon pairs source \textbf{EPS} emits a Bell state via SPDC. Alice's half of the Bell state (= $\ket{\psi_A}$) is transmitted via the quantum channel \textbf{QC\_EPS1} while Bob's half of the Bell state (= $\ket{\psi_B}$) is transmitted via the quantum channel \textbf{QC\_EPS2}. At the same time, Alice also generates the target single photon pulse (= $\ket{\psi_T}$) via the weaklaser \textbf{WL} (again, a laser \textbf{L} along with 2 ND filters (\textbf{NDF\_1} and \textbf{NDF\_2})) and transmits it via the quantum channel \textbf{QC\_WL}. At the receiver end, \textbf{QC\_EPS1} and \textbf{QC\_WL} are connected to the non-polarizing beam splitter \textbf{NPBS}. The lengths of both the quantum channels: \textbf{QC\_EPS1} and \textbf{QC\_WL} are kept the same so as to ensure that Alice's half of the Bell state (= $\ket{\psi_A}$) and the target single photon pulse (= $\ket{\psi_T}$) arrive at nearly the same time at the 2 input ports of the 50:50 \textbf{NPBS}.

The 2 photon pulses with states $\ket{\psi_A}$ and $\ket{\psi_T}$ are superposed at the input of the 50:50 \textbf{NPBS}. It is here that the Bell State Measurement (BSM) setup comes into play. BSM involves the projection of the product (\textit{unentangled}) state of the 2 input photon pulses, say, $\ket{\psi_{pr}} = \ket{\psi_A} \otimes \ket{\psi_T}$, onto the basis of the 4 maximally \textit{entangled} (Bell) states. It is easy to see that this projection is probabilistic. Given any 2 photon state:
\begin{equation}
\ket{\psi_{2p}} = \alpha \ket{00} + \beta \ket{01} + \gamma \ket{10} + \delta \ket{11},
\end{equation}
we can re-express it in terms of the Bell basis as:
\begin{equation*}
\ket{\psi_{2p}} = \frac{\alpha + \delta}{\sqrt{2}} \ket{\phi^+} + \frac{\alpha - \delta}{\sqrt{2}} \ket{\phi^-} + \frac{\beta + \gamma}{\sqrt{2}} \ket{\psi^+} + \frac{\beta - \gamma}{\sqrt{2}} \ket{\psi^-},
\end{equation*}
where, $\ket{\phi^\pm}$ and $\ket{\psi^\pm}$ were defined in (\ref{eqna23}) and (\ref{eqna24}) respectively.

Accordingly, once $\ket{\psi_{pr}}$ is projected onto a Bell state, the state at the output of the \textbf{NPBS} can be determined. The \textbf{NPBS} is connected to 2 polarizing beam splitters \textbf{PBS1} and \textbf{PBS2} via the quantum channels \textbf{QC\_PBS1} and \textbf{QC\_PBS2} attached to its 2 output ports. \textbf{PBS1} will generally transmit a horizontally polarized photon via quantum channel \textbf{QC\_D2} to detector \textbf{D2} and reflect a vertically polarized photon via quantum channel \textbf{QC\_D1} to detector \textbf{D1}. \textbf{PBS2} will generally transmit a horizontally polarized photon via quantum channel \textbf{QC\_D3} to detector \textbf{D3} and reflect a vertically polarized photon via quantum channel \textbf{QC\_D4} to detector \textbf{D4}. Following (\ref{eqnbs19}), (\ref{eqnbs20}), (\ref{eqnbs21}) and (\ref{eqnbs22}) and the Bell state detection analysis described earlier in Section~\ref{BSDA}, we can say that the post-BSM state is:
\begin{itemize}
\item $\ket{\phi^\pm}$: When either \textbf{D1} or \textbf{D2} or \textbf{D3} or \textbf{D4} register 2 $\approx$ simultaneous clicks.
\item $\ket{\psi^+}$: When either \textbf{D1} and \textbf{D2} or \textbf{D3} and \textbf{D4} simultaneously register a click each.
\item $\ket{\psi^-}$: When either \textbf{D2} and \textbf{D4} or \textbf{D1} and \textbf{D3} simultaneously register a click each.
\end{itemize}
Evidently, our linear-optics based \textbf{NPBS} cannot distinguish between $\ket{\phi^+}$ and $\ket{\phi^-}$ and hence, only a partial BSM is possible. Accordingly, we restrict ourselves to the analysis of the cases when only either of $\ket{\psi^+}$ or $\ket{\psi^-}$ is detected. 

Post-BSM, Bob's photon would be projected to a quantum state that depends upon the target state, initially shared Bell state and finally detected Bell state~\cite{b35}. For e.g., if the initially shared Bell state between Alice and Bob was:
\begin{equation}
{\ket{\phi^-}}_{AB} = \frac{1}{\sqrt{2}} ({\ket{0}}_A{\ket{0}}_B - {\ket{1}}_A {\ket{1}}_B),
\label{QTBobeqn1}
\end{equation}
where, the subscripts $A$ and $B$ are used to demarcate Alice's and Bob's halves of the shared pair respectively.
And if post-BSM, photon $A$ and photon $T$ (the target photon) are projected to and detected as being in the Bell state:
\begin{equation}
{\ket{\psi^+}}_{AT} = \frac{1}{\sqrt{2}} ({\ket{0}}_A{\ket{1}}_T + {\ket{1}}_A {\ket{0}}_T),
\label{QTBobeqn2}
\end{equation}
post-BSM, the quantum state of Bob's photon $B$ would become:
\begin{equation}
\ket{\psi_B} = ZX \ket{\psi_T}.
\label{QTBobeqn3}
\end{equation}
This can be understood as follows. Since we observed photons $A$ and $T$ in the state ${\ket{\psi^+}}_{AT}$, we know that photon $A$ is in a state orthogonal to that of photon $T$, i.e., $\ket{\psi_A} = X \ket{\psi_T}$. However, since photons $A$ and $B$ are entangled in the state ${\ket{\phi^-}}_{AB}$, we already know that photon $B$ is in a phase-flipped version of the state of photon $A$, i.e., $\ket{\psi_B} = Z \ket{\psi_A}$. Accordingly, it is evident that $\ket{\psi_B} = ZX \ket{\psi_T}$ as attested in (\ref{QTBobeqn3}).

All the possible combinations and the corresponding state of Bob's photon post-BSM can be similarly derived. The results are shown in Table~\ref{tabQTBob}.

\begin{table}[htbp]
\begin{center}
\caption{Quantum state of Bob's half of the shared Bell state after BSM}
\begin{tabular}{|c|c|c|}
\hline
\textbf{Initially Shared}&\textbf{Detected}&\textbf{Quantum State}\\
\textbf{Bell State}&\textbf{Bell State}&\textbf{of Bob's Photon}\\
\textbf{(Before BSM)}&\textbf{(Via BSM)}&\textbf{(After BSM)}\\
\hline
\multirow{2}{1 cm}{\centering $\ket{\phi^+}$}&$\ket{\psi^+}$&$X \ket{\psi_T}$\\
\cline{2-3}
&$\ket{\psi^-}$&$XZ \ket{\psi_T}$\\
\hline
\multirow{2}{1 cm}{\centering $\ket{\phi^-}$}&$\ket{\psi^+}$&$ZX \ket{\psi_T}$\\
\cline{2-3}
&$\ket{\psi^-}$&$ZXZ \ket{\psi_T}$\\
\hline
\multirow{2}{1 cm}{\centering $\ket{\psi^+}$}&$\ket{\psi^+}$&$\ket{\psi_T}$\\
\cline{2-3}
&$\ket{\psi^-}$&$Z \ket{\psi_T}$\\
\hline
\multirow{2}{1 cm}{\centering $\ket{\psi^-}$}&$\ket{\psi^+}$&$XZX \ket{\psi_T}$\\
\cline{2-3}
&$\ket{\psi^-}$&$XZXZ \ket{\psi_T}$\\
\hline
\end{tabular}
\label{tabQTBob}
\end{center}
\end{table}

Bob then measures his photon's quantum state in the \emph{Z} Basis with the help of the polarizing beam splitter \textbf{PBS\_T} which will generally transmit a horizontally polarized photon via quantum channel \textbf{QC\_DT2} to detector \textbf{DT2} and reflect a vertically polarized photon via quantum channel \textbf{QC\_DT1} to detector \textbf{DT1}. Subsequently, he uses the BSM result to (if required) operate on his state so as to \textit{theoretically} make $\ket{\psi_B} = \ket{\psi_T}$ (again see Table~\ref{tabQTBob}). The BSM result is sent by Alice to Bob via the classical channel \textbf{CC}. A bit = $0$ sent by Alice to Bob indicates that Alice has detected the state $\ket{\psi^+}$ after BSM while a bit = $1$ sent by her indicates the detection of the state $\ket{\psi^-}$ after BSM. Post such (if required) corrections, $\ket{\psi_B}$ can finally be \textit{practically} compared with the initially sent $\ket{\psi_T}$ so as to ascertain the successfulness of the entire quantum teleportation network.

\subsubsection{Results}
\label{QTres}
\begin{figure}[htbp]
\begin{center}
\includegraphics[scale = 0.55]{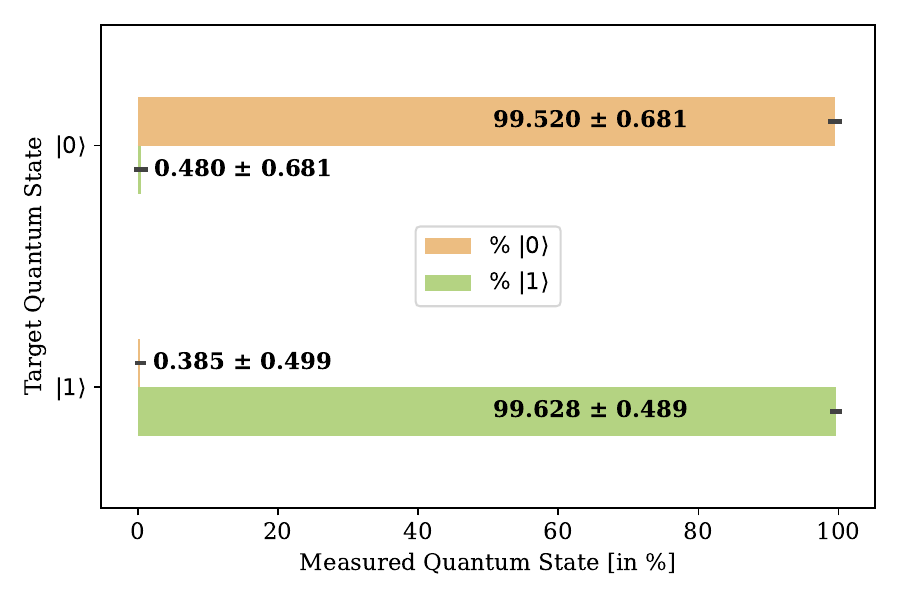}
\end{center}
\caption{Post-correction measured quantum states resulting from QT simulated using the setup illustrated in Fig.~\ref{figQT}}
\label{QTres_fig}
\end{figure}

In line with the details mentioned in Section~\ref{QTimpl}, we simulated QT using the experimental setup shown in Fig.~\ref{figQT}. The results were averaged across 10 runs for each of the two target quantum states. Teleportation was attempted 2500 times in each run. On average, it was found that $\approx$ 104 qubits and 106 qubits (corresponding to the 2 target quantum states $\ket{0}$ and $\ket{1}$ respectively) were successfully transmitted by Alice and detected by Bob. Fig.~\ref{QTres_fig} depicts the averages (bar heights) and standard deviations (errorbars) of the percentages of the measured quantum states (post any necessary corrections) for each of the two target quantum states. 

A maximum error of $\approx$ 2\% (minimum fidelity $\approx$ 98\%) was observed in teleportation. This trend is in line with previous quantum teleportation experiments~\cite{QT1, QT2} and again attests to the accurate consideration of intra and inter-component errors such as finite efficiencies, transmittances, polarization extinction ratios, detector dark counts, and different types of noise in ShaNQar. 

\section{Related Work}

Several Quantum Network Simulators (QNSs) have been developed by numerous R\&D teams across the globe. QuNetSim (\textbf{Qu}antum \textbf{Net}work \textbf{Sim}ulator)~\cite{b36}, SimQN~\cite{b37}, and SimulaQron~\cite{b38} do not model any realistic physical components used in quantum networks and do not implement any timing control and synchronization of components. QuNetSim caters \emph{majorly} to educational needs by allowing users to start their journey of learning about quantum networking by simulating certain quantum networking protocols to gain a better understanding, albeit in an unrealistic way~\cite{b39}. SimQN is an enhanced version of QuNetSim in that it can simulate large-scale quantum networks. However, as stated by its authors, not all its functions are well-tested as yet~\cite{b40}. SimulaQron, on the other hand, specifically focusses on application development and is only intended to be a tool for software engineers for testing basic algorithmic implementations of protocols before possibly deploying them on a real-life quantum internet~\cite{b38}. 

Another QNS, SQUANCH (\textbf{S}imulator for \textbf{Qua}ntum \textbf{N}etworks and \textbf{Ch}annels)~\cite{b41}, although parallelized, uses only the mathematical formalism of QCQI so as to realize (only \emph{theoretically}) accurate simulations of quantum networks. However, like the aforementioned 2 simulators, it also lacks in physical layer modelling and timing control and synchronisation which are crucial for building a QNS which is practically useful.

QuISP (\textbf{Qu}antum \textbf{I}nternet \textbf{S}imulation \textbf{P}ackage)~\cite{b42} only tracks the error state of qubits by using an error probability vector and not their actual quantum state. Therefore, it is neither suitable for modelling the physical layer, i.e., the quantum hardware, nor any quantum algorithms relying on the actual quantum state of a qubit. Additionally, it is only capable of tracking a limited number of errors, \emph{viz.}, Pauli \emph{X}, \emph{Y}, and \emph{Z} errors, relaxation and excitation errors, and photon loss. Important errors such as decoherence and dephasing have not been considered. Effectively, it can only be used for affecting a rather high-level approximation of error state evolution.

Two other recent QNSs, NetSquid (\textbf{Net}work \textbf{S}imulator for \textbf{Qu}antum \textbf{I}nformation using \textbf{D}iscrete Events)~\cite{b43} and SeQUeNCe (\textbf{S}imulator of \textbf{Qu}antum \textbf{N}etwork \textbf{C}ommunication)~\cite{b44, b45, b45a, b45b} are more relevant for comparison with our QNS, ShaNQar, since they implement timing control and synchronization and \emph{to some extent}, model relevant physical components allowing for a hardware level simulation of quantum communication protocols. We will now compare ShaNQar with the corresponding components, features, etc. in NetSquid and SeQUeNCe based on the quantum communication modalities implemented by ShaNQar. 

All the 3 simulators support both the ket and density matrix formalisms for the storage of a quantum state. However, both NetSquid and SeQUeNCe use their own simulation kernels which have simulation time resolution limits of nanoseconds and picoseconds respectively. This essentially leads to the neglect of any events taking place at a time scale faster than either of these, say, for e.g., the femtosecond scale. Thus, this definitely acts as an impediment in the accurate timing control and synchronization of ultrafast devices such as femtosecond lasers. ShaNQar overcomes this limitation by using SimPy~\cite{b12}. SimPy has \emph{virtually}, no time resolution limit. We emphasize on `virtually' since the only time resolution limit imposed by ShaNQar is that by the underlying hardware on which it is run, such as for instance, $10^{-308}$s, (as dictated by the IEEE-754 binary64 format~\cite{b46} which is followed by most of the modern day computers across the globe) which in any case is far more than sufficient for any accurate simulation. 

Secondly, while all three simulators are customizable, both NetSquid and SeQUeNCe only model a limited number of components and further, only consider a limited number of parameters corresponding to those components. This imposes a further constraint on the user since important components such as attenuators like ND filters, waveplates, etc. have not been modelled at all and parameters such as temporal width accrued due to dispersion, detection timing jitter, etc. have not been considered. ShaNQar includes a large number of tunable, i.e., user configurable parameters and also models physical components that have not been previously modelled by either of them. Additionally, the way in which certain parameters have been considered by SeQUeNCe is not entirely correct (It is not possible for us to ascertain the correctness of NetSquid however since it is not open-source.). For instance, in its modelling of SPDs, for deciding on the final detection events, SeQUeNCe simply considers the earliest set of detection events and does not consider any cutoff timings for determining the final detection events~\cite{b47}. This is clearly incorrect since dark counts triggering false photon counts much before or much after the cutoffs cannot possibly correspond to a photon detection event in any case. ShaNQar overcomes this inaccuracy in modelling by carefully analysing detection events with the help of cutoff times (as already detailed in Sections~\ref{SPDA} and~\ref{BSDA}). ShaNQar thus provides for increased variability and versatility while at the same time ensuring reliability and accuracy. 

Finally, for an easier code-based simulation at the user's end, all our components allow for the establishment of direct internal linkages. Only the relevant function of the starting source component has to be explicitly invoked by the user; functions of other connected components are automatically invoked and executed. This allows for a kind of \emph{plug and play} feature which is absent in both NetSquid and SeQUeNCe since they require the user to explicitly invoke the functions for all the connected components as well for customized protocols/applications, i.e, for which they have not already defined separate classes with all the relevant code written beforehand.  

Thus, based on what currently exists in ShaNQar, it is more accurate, versatile, and user-friendly than the existing QNSs.

\section{Conclusion}

In this work, we have thus presented ShaNQar (Simulator of Network Quantique): a modular, customizable, and versatile QNS comprising models of relevant physical components with multiple parameters for variability and versatility. It has (virtually) no simulation time resolution limit and is a `plug and play' simulator since it only requires connecting the components in the network and invoking the source's relevant function to run a simulation of the entire quantum network. We successfully simulated QKD and QT setups and achieved 1 - 4\% QBERs and KGRs in the range of 0.1 Kb/s to 0.1 Mb/s in QKD and a maximum error of 2\% in QT. Our results are in line with similar real-life experiments conducted previously and thus, serve as a validation of accurate and reliable modelling of quantum networks by ShaNQar. Evidently, ShaNQar can therefore be used for flexibly simulating quantum networks at scale while triaging potential quantum hardware improvements and optimizing quantum communication protocols.

Future work includes the extension of ShaNQar to other frequently used quantum information encoding schemes such as time-bin and phase, modelling of other important physical components such as quantum memories for simulating complex entanglement management between multiple parties and finally, its parallelization~\cite{b48} so that it can be run faster and more efficiently. 

\begin{acknowledgments}
    The authors thank Dr. Asvija Balasubramanyam, Joint Director, Systems Development and Operational Services (SDOS) Group and his team at the Centre for Development of Advanced Computing (C-DAC), Bengaluru for their help and support during the project's foundational stages. The authors also thank Prof. C.M. Chandrashekhar and his team at the Department of Instrumentation and Applied Physics (IAP), Indian Institute of Science (IISc) Bengaluru and Prof. R. P. Singh at the Quantum Technologies (QuTe) Laboratory, Physical Research Laboratory (PRL), Ahmedabad for their constructive suggestions and feedback which helped improve ShaNQar.
\end{acknowledgments}

\end{document}